\documentclass[superscriptaddress,prb,super=false,
showpacs,floatfix,fleqn,longbibliography]{revtex4-1}

\UseRawInputEncoding
\setcitestyle{numbers,square}

\usepackage[dvips]{graphicx}
\usepackage{epsfig} 
\usepackage{subfig}
\usepackage{color}
\usepackage[normalem]{ulem}
\usepackage{amsmath,amsfonts,amssymb,bm}
\setcitestyle{numbers,square}
\UseRawInputEncoding

\begin{document}
\title{Two stage decoherence  of optical phonons   in long  oligomers}
\author{Alexander L. Burin,  Igor V. Rubtsov} 
\email[]{aburin@tulane.edu}
\affiliation{Department of Chemistry, Tulane University, New
Orleans, LA 70118, USA}
\date{\today}
\begin{abstract}
Intramolecular energy transport is generally responsible for  chemical energy balance in molecular systems.  The transport  is   fast and efficient if energy is transferred by optical phonons in periodic oligomers, but its  efficiently is limited by decoherence emerging due to anharmonic  interactions with acoustic phonons.  We show   that  in the most common case of the optical phonon band being narrower than  the acoustic bands   decoherence takes place in two stages.  The faster   stage involves  optical phonon  multiple forward scattering due to absorption and emission of transverse acoustic phonons, i. e. collective bending modes with a quadratic spectrum; the transport remains
ballistic and the speed can be altered.   The  subsequent slower stage involves phonon   backscattering in multiphonon processes involving two or more acostic phonons resulting is a switch to diffusive transport.   If the initially excited optical phonon  possesses a relatively small group velocity,  then its equilibration in the first stage is accompanied by its acceleration due to its transitions  to states  propagating faster. This theoretical expectation  is consistent with the recent measurements  of  optical phonon transport in alkane chains, accelerating  with increasing the chain length.   
\end{abstract}

\maketitle



\section{Introduction}
\label{sec:Intr}

Intramolecular vibrational energy transport is responsible for  chemical energy balance \cite{AbeScience07,
15LeitnerReview,
PandeyLeitner2016ThermSign,Leitner2019EnTransBiomol,
Karmakar20VibrMBL} controlling reaction kinetics   \cite{Zaragoza2023VibrEnHtunnProt,
Jiang2020NuclReorg} and it has  potential applications in  several fields of modern technology  spanning from sustainable energy to biomedicine to thermal management \cite{Zhu19AdvMatVibrEnTr}.  It  can be fast and efficient even compared to that in metals because it emerges due to covalent bonds, that are the strongest chemical bonds existing in nature  \cite{Abe03,DlottScience07,
Asegun08Polyethilenetransport,15LeitnerReview,SegalReview2016}.  Although thermal conductivity of molecules at room temperature or below is determined substantially by low energy acoustic phonons \cite{SegalNitzan03,SegalReview2016,
Abe2020AtomicCurrInterf,
Dvira2022QuantTranspRev}, high energy optical phonons can be in charge for chemical energy transport. \cite{ab19IgorReview,Cui2019SingleMoleculeThermCond,
Leitner2019EnTransBiomol}.  Indeed,   typical chemical energy,   exceeding phonon energy by over an order of magnitude, can be  released easier to optical phonons rather than to acoustic phonons possessing considerably  smaller energy.  
Since  their  propagation speeds are comparable \cite{ab2023Cherenkov}, optical phonons can transfer energy more efficiently compared to the acoustic ones.    

The fast and efficient energy transport by optical phonons has been  demonstrated in numerous experiments \cite{Rubtsov12Acc,ab15AccountsIgor,ab19IgorReview,
Rubtsov2012pphynilultrfast,
Rubtsov2009Accounts2DIR,ab15ballistictranspexp,
ab19layla}.  If an optical phonon is initiated within the band of a periodic oligomer chain it can propagate along the chain ballistically due to normal mode delocalization \cite{ab19IgorReview}.  
This is the most efficient energy transport realization, which  is limited by decoherence and relaxation of propagating phonons induced by their anharmonic interactions with the other intramolecular vibrations and the environment  \cite{ab2023Cherenkov}.  Here we consider  intramolecular interactions since the coupling to the environment is much weaker \cite{Karmakar20VibrMBL}.

Following earlier work \cite{ab15jcp} we specify decoherence as  optical phonon transitions  holding them   within the same  band, while relaxation  irreversibly removes optical phonons away from  the band.  Decoherence occurs inevitably due to interaction with low frequency acoustic phonons, while relaxation requires involvement of other optical phonons and is usually slower \cite{ab15jcp,ab15JPCExpDec,ab16jpcPEGs}. Consequently, decoherence is the first possible source  of  ballistic transport breakdown to consider.  In the present work {\it we focus on optical phonon decoherence due to their anharmonic interaction with transverse acoustic phonons.}

Decoherence induced by interaction with longitudinal or torsional acoustic phonons possessing a sound like spectrum was considered in the earlier work \cite{ab2023Cherenkov}. It was shown that it is substantially suppressed in the most common situation of a forbidden Cherenkov's like emission (or absorption), where the speed of sound $c$ exceeds the maximum optical phonon velocity $v_{max}$.  In this regime  only relatively slow high order anharmonic processes (forth and higher) results in phonon scattering in contrast with the regime of a violated Cherenkov's constraint $v_{max}>c$. 

Decoherence due to transverse acoustic phonons representing collective bending modes needs special consideration because these phonons possess a quadratic spectrum at small wavevectors ($\omega=Aq^2$, see e. g.  Ref.  \cite{ab20Transv}). For this spectrum there always exist acoustic phonons with a  small velocity violating  the Cherenkov's constraint,   so absorption and emission are always allowed.  However,  if an optical phonon band is narrower   than that for transverse acoustic phonons, then the absorption or emission of transverse phonon leads to a small change in the optical phonon wavevector compared to this wavevector.    Consequently,    absorption or emission do not modify  the direction of the optical phonon propagation leading only to its forward scattering. Backscattering requires  multiphonon processes occurring much slower compared to single phonon absorption or emission.    
Thus  phonon  relaxation  occurs   in two (fast and slow) stages. The fast stage taking  around $1$-$10$ ps  involves an incomplete  equilibration of the optical phonon within the states of the band with group velocities oriented towards its  initial direction.  At that stage the energy  propagates with a nearly constant velocity similarly to the ballistic transport regime.  The second stage, involving phonon backscattering,   emerges after orders of magnitude longer time.  Only at that stage the transport mechanism changes from the ballistic transport to the  diffusive transport.  

It is noticeable that if the optical phonon  velocity is small in the initial state compared to a velocity  typical value,  then the phonon propagation speed can increase with time during the first stage. This increase can be used to interpret the observations of Ref. \cite{2024IgorBallTr}  in alkane chains, where an increase of the optical phonon transport velocity has been discovered with increasing the  chain length. 
 
 In the opposite case of a narrow transverse phonon band, any absorption or emission of transverse phonons overturns the propagating phonon backwards leading to a rapid ballistic transport breakdown similarly to Ref. \cite{ab2023Cherenkov}.  
 
 The paper is organized as follows. In Section \ref{sec:Mod} the model of optical and transverse acoustic phonons and their anharmonic interactions is introduced.  In Section  \ref{sec:EnMomCons} the phonon decoherence  is  considered in terms of elementary processes of transverse phonon emission, absorption or scattering.  The rates of these processes  are  estimated using Fermi golden rule. Our estimates are found to be  consistent with the numerical simulations   reported in Section  \ref{sec:num} performed using an accurate quantum mechanical  \cite{ab19FPU,ab2023Cherenkov} and semiclassical  \cite{Senthilkumar2005DNASemicl} approaches.         In Section  \ref{sec:exp} we examine numerically  the increase of the transport speed of optical phonon with the time assuming its  small initial  velocity  and discuss the possible connection of this result to the recent experiments  in alkane chains \cite{2024IgorBallTr}.   Finally, the conclusions are  formulated in Section \ref{sec:concl}.   


\section{Model}
\label{sec:Mod}

Here we introduce the model for optical phonons in a periodic chain interacting with transverse acoustic phonons.   Our consideration is limited to a single optical phonon interacting with acoustic phonons, which is consistent with typical experimental conditions \cite{ab19IgorReview}. 

\subsection{Boundary conditions and normal modes}
\label{sub:ModGen}

The molecule is represented by a circular periodic  chain of $N$ identical sites (unit cells) separated by distance $a$. Although real chains are not circular, this model is still relevant since  the propagation of phonon should not be sensitive to boundaries if the chain length $L=Na$ exceeds the coherence length of a pure ballistic propagation of phonons.  In this regime the boundary conditions are not  significant.  In the opposite regime where the transport remains ballistic at distances exceeding molecular length   the decoheherence is irrelevant.  The use of periodic model  simplifies numerical studies because of the quasi-wavevector  conservation    \cite{ab19FPU} that permits us to investigate numerically longer chains than with other boundary conditions.  Normal modes of the periodic chain for any specific band  can be enumerated by integer numbers  $n$  and characterized by a wavevector 
\begin{eqnarray}
k=\frac{2\pi n}{L}, ~ L=Na, ~ n=\begin{cases}
      -\frac{N-1}{2},...\frac{N-1}{2},  & \text{if $N$ is odd,}\\
      -\frac{N}{2}+1,...\frac{N}{2} & \text{if $N$ is even}.
    \end{cases} 
\label{eq:wvvector}
\end{eqnarray}
Wavevectors are chosen to make phonon states periodic since their site dependence is determined by the exponent $e^{iqan}$.   

\subsection{Optical phonons} 
\label{sub:modopt}

We describe the optical phonon band using a standard Hamiltonian \cite{ab2023Cherenkov} 
\begin{eqnarray}
\widehat{H}_{\rm opt}=\sum_{k}\hbar\omega_{\rm opt}(k)\widehat{a}_{k}^{\dagger}\widehat{a}_{k},
\label{eq:Ham0pt}
\end{eqnarray}
where the bosonic operators $\widehat{a}_k$ ($\widehat{a}_k^{\dagger}$)  represent annihilation (creation) operators of optical phonons characterized by the wavevector $k$ Eq. (\ref{eq:wvvector}) and $\omega_{\rm opt}(k)$ stands for  the wavevector dependent phonon frequency  It is a periodic function of the  wavevector  with the period $2\pi/a$ and it is an even function  because of the time reversal invariance We model the phonon spectrum with the simplest periodic function 
\begin{eqnarray}
\omega_{\rm opt}(k) = \omega_{0}+ \frac{\Delta_{\rm opt}(1-\cos(ka))}{2}.,  
\label{eq:OptPhNN}
\end{eqnarray}
corresponding to the nearest neighbor interaction of site vibrations in the coordinate representation. \cite{ab2023Cherenkov}  Here $\omega_{0}$ is the optical phonon bandgap $0<\Delta_{\rm opt} \ll \omega_{0}$ and $\Delta_{\rm opt}$ is the bandwidth. We consider  direct bands with $\Delta_{\rm opt}>0$.  In all calculations below we assume $\omega_{0}=1000$ cm$^{-1}$ and $\Delta_{\rm opt}=100$ cm$^{-1}$ similarly to Ref. \cite{ab2023Cherenkov},  which is quite consistent with optical phonon energy band properties for organic polymer  chains \cite{ab15jpc}.  Generalization to a more complicated spectrum is straightforward.

The optical phonon transport is defined by the phonon current that   reads 
 \begin{eqnarray}
\widehat{J}=\sum_{k}v_{\rm opt}(k)\widehat{a}_{k}^{\dagger}\widehat{a}_{k}, ~~v_{\rm opt}(k) =\frac{\partial\omega_{\rm opt}}{\partial k}  =\frac{a\Delta}{2}\sin(ka),
\label{eq:EnFlux}
\end{eqnarray}
where $v_{\rm opt}(k)$ is  velocity  of the phonon with the given wavevector $k$, which coincides with  the group velocity for the same wavevector.  
This average  current is evaluated numerically  in Sec.  \ref{sec:num} to characterize the energy transport  and its suppression  by decoherence.

\subsection{Transverse acoustic phonons} 
\label{sub:modtr}

Transverse acoustic phonons, i. e. collective bending modes,  are discarded in most of considerations of thermal energy transport possibly  because  they possess a vanishing group velocity $v(q) \propto q$ in the long wavelength limit\cite{landau1986theory,PolimVibr1994Book,
Chico2006VibrCarbNanot,
NanoTubeVibr20071,ab20Transv,ab15jcp}, while longitudinal phonons possess a constant speed of sound.  However, they provide a more significant  source of dissipation as a sub Ohmic bath \cite{ab20Transv}. so they can be significant for optical phonon decoherence.   


The Hamiltonian of transverse acoustic phonons for each band $\mu=1,2$ corresponding   to two possible directions of transverse displacement can be expressed as 
\begin{eqnarray}
\widehat{H}_{\rm tr}=\sum_{q}\hbar\omega_{{\rm tr}\mu}(q)\widehat{b}_{q\mu}^{\dagger}\widehat{b}_{q\mu},  
\label{eq:HamTr}
\end{eqnarray}
where bosonic operators $\widehat{b}_{q\mu}$ ($\widehat{b}_{q\mu}^{\dagger}$)  represent annihilation (creation) operators of transverse phonons within the normal modes characterized by the operator subscript indices.  

Transverse phonon frequencies are periodic functions of the wavevector, approaching $0$ for $k\rightarrow 0$.  We consider the simplest possible model for their spectrum, that is similar to the one for optical phonons Eq. (\ref{eq:OptPhNN}) but with the zero band gap  
\begin{eqnarray}
\omega_{\rm tr}^{\mu}(q) = \frac{\Delta_{{\rm tr}\mu} (1-\cos(qa))}{2},  
\label{eq:TrNN}
\end{eqnarray}
where $\Delta_{{\rm tr}\mu}$ is the transverse acoustic phonon  bandwidth for the specific branch $\mu$.  The spectrum in Eq. (\ref{eq:TrNN}) is originated from a potential energy determined by  local "bendings" $\epsilon_{i,\mu}=(u_{i-1,\mu}-2u_{i,\mu}+u_{i+1,\mu})/a$ in the form  $U_{bend}^{\mu}=\frac{A_{\mu}}{2}\sum_{i}\epsilon_{i,\mu}^2$, where the parameter $A_{\mu}$ is defined as $A_{\mu}=Ma^2\Delta_{{\rm tr}\mu}^2/4$ and $M$ is the mass of the elementary cell. 

If the molecule is immersed into a solvent, then the  acoustic  phonon spectrum might acquire a small gap, because the solvent violates the translational invariance \cite{Kittel2004}  for atoms belonging to the molecule.    Since this gap is determined by the interaction with the environment that is much weaker compared to intramolecular interaction, we ignore it. 

We limit the further  consideration to the interaction of optical phonons with a single transverse acoustic band, since the addition of the second band will not modify results qualitatively, while the generalization to two bands is quite straightforward.   Consequently, we skip the earlier introduced index $\mu$ everywhere  for the single transverse band considered below. 

Similarly to Ref.  \cite{ab2023Cherenkov},   the relationship of acoustic and optical phonon  bandwidths is a critically important parameter. We introduce it as a separate ratio parameter $r$ defined as 
\begin{eqnarray}
r=\frac{\Delta_{\rm tr}}{\Delta_{\rm opt}}. 
\label{eq:RatPar}
\end{eqnarray}
Acoustic bands are wider than optical bands since they are determined by  direct atomic interactions through covalent bonds so usually,  one has $r>1$ (see e. g. \cite{ab15JPCExpDec}).  

\subsection{Anharmonic interaction of optical phonons with transverse phonons} 
\label{sub:opttrint}

Anharmonic interactions are weaker compared to harmonic ones by the ratio  of the typical vibration amplitude and interatomic distance represented  by the dimensionless parameter $\eta \sim 0.1$ (see Ref. \cite{ab2023Cherenkov} and Eq.  (\ref{eq:AnhIntC}) below). The effective strength of  the $k^{th}$ order anharmonic interaction is expressed by the parameter $\eta^{k-2}$ since harmonic interaction emerges in the second order with respect to the parameter $\eta$.  
We consider  the strongest  third order anharmonic interaction similarly to Ref. \cite{Borrelli2020BallAnhInit}, where it was considered for the wavepacket initiation. In the third  order the only processes capable to conserve energy are determined by  absorption or emission of transverse  phonons by optical phonons, which can be expressed   as  
\begin{eqnarray}
\widehat{V}_{\rm anh}=\frac{\hbar}{\sqrt{N}}\sum_{k, q}^{'}\left[ V(q,k)\widehat{a}_{k}^{\dagger}\widehat{a}_{k+q}\widehat{b}_{q}^{\dagger}  + H. C.\right], 
\label{eq:AnhIntGen}
\end{eqnarray}
where $\sum^{'}$ means that we use the periodic extension of the wavevector $k+q$ replacing it with $k+q \pm 2\pi/a$ if it is outside the domain of the wavevectors  $(-\pi/a, \pi/a)$.  $V(k, q)$ is the interaction constant of optical and transverse acoustic phonons; one has $V(q,k) \propto q$ for $q\rightarrow 0$ \cite{ab20Transv}. 

For the future consideration we take the interaction in the simplest form corresponding to the coordinate representation 
\begin{eqnarray}
\widehat{V}_{\rm loc}=\frac{\hbar V_{0}}{a}\sum_{i=1}^{N}\widehat{a}_{i}^{\dagger}\widehat{a}_{i}(\widehat{u}_{i+1}-2\widehat{u}_{i}+\widehat{u}_{i-1}),
\label{eq:ModAnh}
\end{eqnarray}
where $V_{0} \sim \Delta_{\rm opt,tr}$ is the interaction constant
and operators  $\widehat{u}_{i}$ stand for  the transverse displacements  at the site $i$ \cite{ab2023Cherenkov} interacting with the local density optical phonons ($\widehat{a}_{i}^{\dagger}\widehat{a}_{i}$).   For this specific  model  the interaction constants $V(q, k)$ can be expressed as 
\begin{eqnarray}
V(q, k) 
=2 V_{3}\left|\sin\left(\frac{qa}{2}\right)\right|, ~ V_{3}=\eta V_{0},  ~ \eta=\sqrt{\frac{2\hbar }{Ma^2\Delta_{\rm tr}}} \sim 0.1,
\label{eq:AnhIntC}
\end{eqnarray} 
This definition is used in all considerations below.  The coupling constant $V_{3}$ represents  a  characteristic strength of anharmonic interactions that should be smaller compared to the harmonic interactions by the small factor $\eta$.

\subsection{Full Hamiltonian}
\label{sub:Ham}

The full Hamiltonian of the system  includes Hamiltonians of optical Eq. (\ref{eq:Ham0pt}) and transverse  Eq. (\ref{eq:HamTr}) phonons and their interaction Eq. (\ref{eq:AnhIntGen}) as rewritten below for a reader convenience 
\begin{eqnarray}
\widehat{H}_{\rm tot}=\sum_{k}\hbar\left(\omega_{0}+\Delta_{\rm opt}\frac{1-\cos(ka)}{2}\right)\widehat{a}_{k}^{\dagger}\widehat{a}_{k}+\sum_{k}\hbar\Delta_{\rm tr}\frac{1-\cos(ka)}{2}\widehat{b}_{k}^{\dagger}\widehat{b}_{k}
\nonumber\\
+\frac{2\hbar V_{3}}{\sqrt{N}}\sum_{q,k}^{'}\left[ \sin(qa/2)\widehat{a}_{k}^{\dagger}\widehat{a}_{k+q}\widehat{b}_{q}^{\dagger}  + H. C.\right], 
\label{eq:FullH}
\end{eqnarray}
using Eqs. (\ref{eq:OptPhNN}) and (\ref{eq:TrNN}) for optical and transverse phonon frequencies, respectively, and Eq. (\ref{eq:AnhIntC}) for anharmonic interactions. 

\subsection{Higher order anharmonic interactions}

The model under consideration includes only   the third order anharmonic interactions Eq. (\ref{eq:AnhIntGen}).  However,  higher order  interactions   are generated  by the third order interactions in a pertubation theory  \cite{ab19FPU}. Therefore, all decoherence channels originated from higher order interactions  are present  within our model Eq. (\ref{eq:FullH}) in the proper order in the small parameter $\eta$ Eq. (\ref{eq:AnhIntC}).  For the future consideration we need the fourth order interaction,because it is   responsible for the optical phonon backscattering in the case of a narrow optical phonon band ($r>1$).   The fourth order interaction is very important because it contains the resonant scattering  of optical and acoustic phonon with opposite momentum expressed for example by the terms like  $a_{-k}^{\dagger}b_{k}^{\dagger}a_{k}b_{-k}$ overturning the optical phonon current.  

We derive the fourth order  interaction using the  celebrated Schrieffer - Wolff transformation   \cite{SchriefferWolff66} eliminating off-resonant third order anharmonic interactions to generate the resonant interaction  in the fourth order.  For the system Hamiltonian separated into the harmonic part $\widehat{H}_{0}$ and the perturbation $\widehat{V}$ (anharmonic interactions) we  introduce the unitary transformation of the Hamiltonian as $\widehat{H} \rightarrow e^{\widehat{S}}\widehat{H}e^{-\widehat{S}}$ with the anti-Hermitian operator  $\widehat{S}$ defined as $\left[ \widehat{S},\widehat{H}_{0}\right]=-\widehat{V}$. Then the expansion of the exponents in the power series in $\widehat{S}$ results in a disappearance of the off-resonant part of the perturbation $\widehat{V}$  (the perturbation expansion is not applicable to a resonant part) with a simultaneous generation of the fourth order interaction in the form \cite{SchriefferWolff66} 
\begin{eqnarray}
\widehat{H}\approx \widehat{H}_{0} + \widehat{V}_{\rm res}+\frac{1}{2}\left[\widehat{S},\widehat{V}_{\rm offres}\right].
\label{eq:SchrWollf}
\end{eqnarray}
where the anharmonic interaction $\widehat{V}$ is split into resonant $\widehat{V}_{\rm res}$ and off-resonant $\widehat{V}_{\rm offres}$ parts. 

For the specific Hamiltonian in Eq.  (\ref{eq:FullH}) one can express the matrix $\widehat{S}$ in the form 
\begin{eqnarray}
\widehat{S}=-\frac{1}{\sqrt{N}}\sum^{'}_{q,k}\left[ V(q, k)\frac{\widehat{a}_{k}^{\dagger}\widehat{a}_{k+q}\widehat{b}_{q}^{\dagger}}{\omega_{\rm opt}(k)-\omega_{\rm opt}(k+q)-\omega_{\rm tr}(q)}  - H. C.\right].  
\label{eq:SMatr}
\end{eqnarray}
The modified summation  $\sum^{'}$ means omitting resonant terms  with $|\omega_{\rm opt}(k)-\omega_{\rm opt}(k+q)-\omega_{\rm tr}(q)|\leq  \eta^2 \Delta_{\rm opt}$  to make both the Schrieffer-Wolff 
perturbation  theory and the Fermi golden rule for the third order anharmonic interaction applicable,  and using the proper definition  of the wavevector sum  $k+q$ shifted by $\pm 2\pi/a$ if necessary., to match the wavevector domain $(-\pi/a, \pi/a)$ Eq. (\ref{eq:wvvector}).   

The induced fourth order anharmonic interaction can be evaluated using Eq. (\ref{eq:SchrWollf}) as
\begin{eqnarray}
\widehat{V}_{4}=
\nonumber\\
=-\frac{4\hbar V_{3}^2}{N\Delta_{\rm opt}}\sum^{'}_{k,q,p}a_{k}^{\dagger}b_{q}^{\dagger}a_{k-p}b_{q+p}\left[\frac{\sin\left(\frac{q+p}{2}\right)^2\cos\left(k-\frac{p}{2}\right)}{\left(r\sin\left(\frac{q}{2}\right)-\sin\left(k+\frac{q}{2}\right)\right)\left(r\sin\left(\frac{q}{2}\right)-\sin\left(k-p-\frac{q}{2}\right)\right)}+\right.
\nonumber\\
\left.+\frac{\sin\left(\frac{q}{2}\right)^2\cos\left(k-\frac{p}{2}\right)}{\left(r\sin\left(\frac{q+p}{2}\right)-\sin\left(k+\frac{q-p}{2}\right)\right)\left(r\sin\left(\frac{q+p}{2}\right)-\sin\left(k-\frac{p+q}{2}\right)\right)}\right].
\label{eq:SchrWlfRes}
\end{eqnarray}

In addition to the induced  fourth order anharmonic interaction  Eq. (\ref{eq:SchrWollf}) there exists the interaction emerging similarly to the third order interaction in Eq. (\ref{eq:AnhIntGen}). This interaction will be also of the second order in the parameter $\eta$ in Eq. (\ref{eq:AnhIntC}) as that in Eq.  (\ref{eq:SchrWlfRes}), so it gives the effect comparable to that of the induced interaction.  Since our consideration is qualitative and does not target the specific molecule, we can skip that interaction.

\section{Optical phonon transitions induced by anharmonic interactions}
\label{sec:EnMomCons} 

Anharmonic interactions of an optical phonon with transverse acoustic phonons Eq. (\ref{eq:AnhIntC}) can lead to  transitions of  this optical phonon between different states emerging in the continuous spectrum for sufficiently long chains ($N\gg 1$).   Such transitions affect a current associated with this phonon.  Particularly, transitions accompanied by the  overturn of the phonon velocity lead to ballistic transport breakdown with its further replacement with the diffusive transport.  

Below we examine the phonon transitions  permitted by the energy and momentum (wavevector) conservation, estimate their rates using Fermi golden rule and consider qualitatively other regimes, not handled by the Fermi golden rule.  In these regimes an  anharmonic interaction is either too strong to be treated as a perturbation or too weak to neglect the discreteness of the system at a finite number of sites $N$.  All regimes are revealed in the numerical studies reported in Sec. \ref{sec:num}.  


\subsection{Emission and absorption}
\label{sub:SingleAbs}

\subsubsection{Energy and momentum conservation}

\begin{figure}
\includegraphics[scale=0.5]{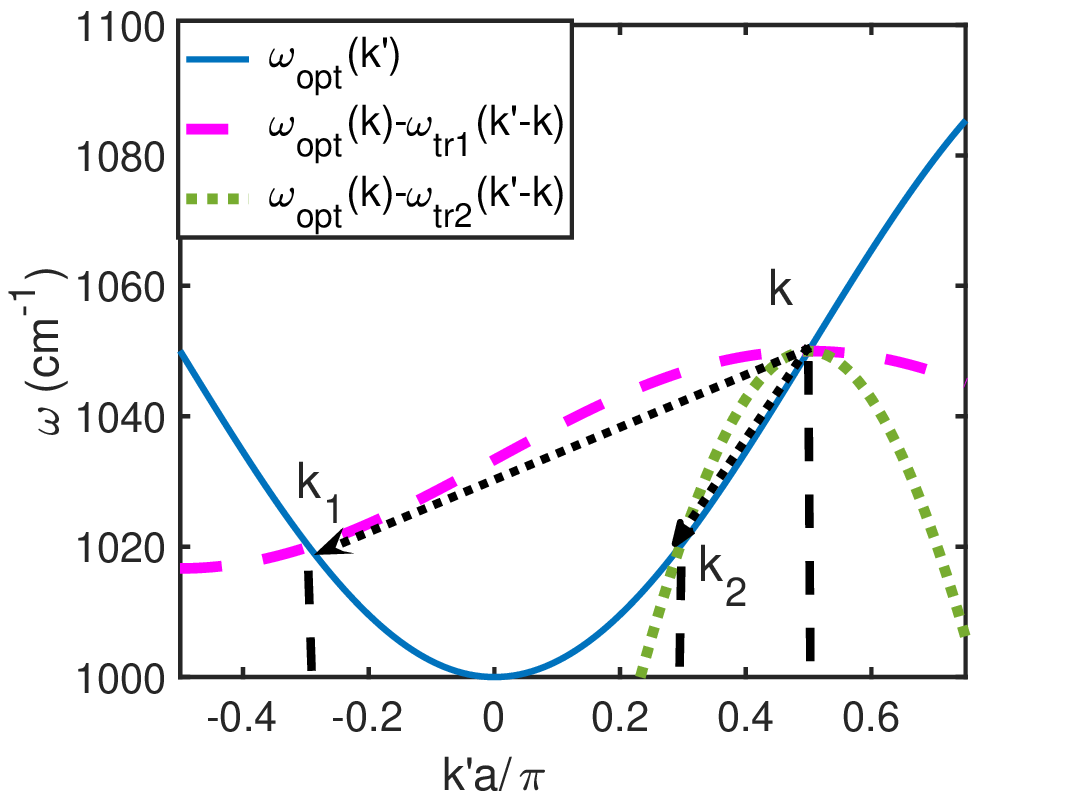} 
\caption{Graphical solution for the emission of transverse phonon Eq. (\ref{eq:EnConsNN}). Initial optical phonon wavevector is $k$ and final is $k'$.  Dashed (magenta) and dotted (green) lines are defined as $y(k')=\omega_{\rm opt}(k)-\omega_{\rm tr}(k-k')$ and used to find the solution of   Eq. (\ref{eq:EnConserv}) for the emitted phonon wavevector $k'$ as the graph  intersection $y(k')=\omega_{\rm opt}(k')$ for  narrow ($\Delta_{\rm tr}=\Delta_{\rm opt}/3$) or wide ($\Delta_{\rm tr}=3\Delta_{\rm opt}$) transverse phonon bands, respectively.   Black arrows illustrate  wavevector modifications after transverse phonon emission.}
\label{fig:Emis}
\end{figure}

Emission or absorption of transverse phonon in a periodic chain emerges with energy and quasi-momentum (wavevector) conservation.  If initially the optical phonon has a wavevector $k$ and after emission it acquires the wavevector $k'$ then the wavevector of emitted transverse phonon is equal to $q=k-k'$ with the accuracy to the inverse lattice period $2\pi/a$.  For  absorption the wavevector of absorbed phonon is  given by $q=k'-k$. In either case the frequency of emitted or absorbed phonon satisfies energy conservation law in the form 
\begin{eqnarray}
\omega_{\rm opt}(k)-\omega_{\rm opt}(k')=\omega_{\rm tr}(k-k')~~( {\rm absorption}),   
\nonumber\\
\omega_{\rm opt}(k')-\omega_{\rm opt}(k)=\omega_{\rm tr}(k'-k)~~ ({\rm emission}).  
\label{eq:EnConserv}
\end{eqnarray} 
These equations are not solvable  for emission of absorption of longitudinal phonons possessing the sound velocity exceeding the maximum group velocity of optical phonons in accord with the Cherenkov's emission criterion  of Ref. \cite{ab2023Cherenkov}.  This is not the case for the transverse phonons as illustrated  in Fig. \ref{fig:Emis} for the modeling phonon spectra Eqs. (\ref{eq:OptPhNN}) and (\ref{eq:TrNN}).  The solution always exists because the group velocity of transverse phonons approaches zero at small wavevector  violating the Cherenkov's constraint. 

Yet the graphical solutions of Eq. (\ref{eq:EnConserv}), expressing the energy conservation law,  are different for different ratios  $r$ of optical and acoustic phonon bandwidths,  as illustrated in Fig.  \ref{fig:Emis} for $r=1/3$ and $r=3$.  Namely, for $r<1$ initial and final wavevector are of opposite signs, so the emission or absorption are accompanied by the overturn of the optical phonon propagation direction, i. e.  backscattering,   while in the opposite case $r>1$ initial and final wavevectors are of the same signs as for  scattering forward.  
This observation is in full accord with  the analytical solution of Eq. (\ref{eq:EnConserv})  for the modeling spectra in Eqs.  (\ref{eq:OptPhNN}) and (\ref{eq:TrNN}),  where we got  (remember, $r=\Delta_{\rm tr}/\Delta_{\rm opt}$)  
\begin{eqnarray}
q =\frac{2}{a}\tan^{-1}\left(\frac{\sin(ka)}{r+\cos(ka)}\right) ~
~ ({\rm emission}),
\nonumber\\
q =\frac{2}{a}\tan^{-1}\left(\frac{\sin(ka)}{-r+\cos(ka)}\right) ~
~({\rm absorption}).
\label{eq:EnConsNN}
\end{eqnarray}

It is clear from both solutions Eq. (\ref{eq:EnConsNN}) that the unity ratio $r=1$ serves as  the crossover regime $q=k$ for emission or $q=\pi/a-k$ for absorption, while for $r>1$ and $k>0$ one has $0<q<k$ for emission and   $0<q<\pi/a-k$ for absorption leaving the wavevector $k'$ positive.  For $r<1$ it inevitably changes direction after both absorption and  emission.  These conclusions remain valid for arbitrarily phonon spectra at sufficiently large  ($r\gg 1$) or small ($r \ll 1$) ratios since for $r\gg 1$ the only small wavevector change leaves phonon energy within the optical band, while for $r \ll 1$ one has $q \approx 2k$ since the narrow band reflects optical phonon backwards almost statically (see also  Fig. \ref{fig:Emis}).

Based on this consideration we can conclude that the emission and absorption of transverse  phonons occur differently depending on the ratio  $r$ of  transverse acoustic and optical phonon bandwidths. If $r>1$, then the change of the wavevector of optical phonon during the emission or absorption is smaller than the wavevector itself. Consequently, single phonon absorption or emission cannot overturn the direction of the phonon propagation. Backscattering requires higher order anharmonic processes considered  below in Sec. \ref{sec:ConsMultPh}. These processes can be orders of magnitude slower   than a single phonon absorption or emission as it is demonstrated in the   numerical simulations reported in Sec. \ref{sec:num}.  Consequently,  a unidirectional  propagation of vibrational energy lasts for a while. Experimentally this  regime is seen  quite similarly to  a ballistic transport since after a phonon equilibration among the states with a certain wavevector direction the transport speed should approach a constant value given by its proper average over those states.  

In the opposite regime of a smaller acoustic phonon bandwidth a single phonon  absorption or emission are accompanied by the change of the direction of phonon propagation. Consequently,   ballistic transport breaks down  after  a single or few absorption or emission events,  which is consistent with our numerical studies reported in Sec.  \ref{sec:num}. 

\subsubsection{Absorption and emission rates}
\label{sec:FGRRates}

Consider the  optical  phonon having the wavevector  $k$. Due to its anharmonic interaction Eq. (\ref{eq:AnhIntGen}) it   can emit  or absorb an acoustic phonon with the wavevector $q$ Eq. (\ref{eq:EnConsNN}) getting  to the new state within the same band, characterized by the wavevectors $k\mp q$  due to the quasi-momentum conservation.  In a first non-vanishing order in a weak  anharmonic interaction the rate of this   transition is given by the Fermi golden rule.  According to this rule emission and absorption rates $W_{\rm em}$, $W_{\rm abs}$ can be expressed as (cf.  Refs.  \cite{Leitner2001ThirdOrdAnh,ab2023Cherenkov})
\begin{eqnarray}
W_{\rm em}=a\int_{-\pi/a}^{\pi/a}dk' V(k-k', k)^2
\nonumber\\
\times\delta\left(\omega_{\rm opt}(k)-\omega_{\rm opt}(k')-\omega_{\rm tr}(k-k')\right)(1+\nu(k-k')), 
\nonumber\\
W_{\rm abs}=a\int_{-\pi/a}^{\pi/a}dk' V(k'-k, k)^2
\nonumber\\
\times\delta\left(\omega_{\rm opt}(k)-\omega_{\rm opt}(k')+\omega_{\rm tr}(k-k')\right)\nu(k-k'), ~\nu(q)=\frac{1}{e^{\frac{\hbar\omega_{\rm tr}(q)}{k_{\rm B}T}}-1},  
\label{eq:FGr}
\end{eqnarray} 
where $\nu(q)$ represents the average number of transverse phonons with the wavevector $q$, defined in Eq. (\ref{eq:EnConsNN}).  

Using the specific phonon spectra and interactions, defined by  Eqs. (\ref{eq:OptPhNN}), (\ref{eq:TrNN}) and  (\ref{eq:AnhIntC}) we evaluate the integral in Eq. (\ref{eq:FGr}) as
\begin{eqnarray}
W_{\rm em}=8\frac{V_{3}^2}{\Delta_{\rm opt}}(1+\nu(q))\frac{\sin(ka)}{r^2+2r\cos(ka)+1}, 
\nonumber\\
W_{\rm abs}=8\frac{V_{3}^2}{\Delta_{\rm opt}}\nu(q)\frac{\sin(ka)}{r^2-2r\cos(ka)+1}, 
\label{eq:FGrSpec}
\end{eqnarray} 
At  high temperatures  where  population numbers can be replaced with their classical expressions  ($\nu(x)=k_{\rm B}T/(\hbar\omega_{\rm tr}(x))$,  and $\nu(x)\gg 1$ we get
\begin{eqnarray}
W_{\rm em}=8\frac{V_{3}^2}{\Delta_{\rm opt}}\frac{k_{\rm B}T}{\hbar\Delta_{\rm tr}\sin(ka)}, ~
W_{\rm abs}=8\frac{V_{3}^2}{\Delta_{\rm opt}}\frac{k_{\rm B}T}{\hbar\Delta_{\rm tr}\sin(ka)}, 
\nonumber\\ 
W_{\rm tot}=W_{\rm abs}+W_{\rm em}=16\frac{V_{3}^2}{\Delta_{\rm opt}}\frac{k_{\rm B}T}{\hbar\Delta_{\rm tr}\sin(ka)}, 
\label{eq:FGrSpechT}
\end{eqnarray} 
where $W_{\rm tot}$ represents the total decay rate of the optical phonon. 

The lifetime $T_{\rm rel}$ of the optical phonon in the given quantum state can be estimated as  the inverse total decay rate
\begin{eqnarray}
T_{\rm rel}=\frac{1}{W_{\rm tot}}. 
\label{eq:LifeTime}
\end{eqnarray}
The inverse lifetime defines the  inelastic width of the energy level as $\hbar/T_{\rm rel}$.  

\subsection{Backscattering for a large bandwidth ratio $r > 1$}
\label{sec:ConsMultPh}


\subsubsection{Backscattering at high temperature.  Fourth order processes.}
\label{sec:BacSchighT}

At a large bandwidth ratio $r>1$ backscattering  can be originated from the phonon scattering expressed by the fourth order anharmonic   interaction Eq. (\ref{eq:SchrWlfRes}), which is the next anharmonic correction to the third order interaction considered above.   There is no constraints for the energy conservation for phonon scattering as can be illustrated for instance by the resonant backscattering of optical and acoustic phonons with opposite wavevectors resulting in their wavevector exchange. 
Such processes require  the presence of excited acoustic phonons  that is possible if the thermal energy $k_{\rm B}T$ is at least comparable or larger than  the transverse bandwidth $\hbar\Delta_{\rm tr}$.  In the opposite case of low temperature $k_{\rm B}T\ll \hbar\Delta_{\rm tr}$ multiphonon emission is needed for the optical phonon back scattering   as shown below in Sec.  \ref{sec:subMultPh}.  

The back scattering  rate can be estimated  using Fermi golden rule with the interaction in Eq. (\ref{eq:SchrWollf}) as a perturbation. Because of the large number of scattering outcomes we do not attempt to calculate a specific relaxation rate, but instead consider its dependence on the anharmonic coupling $V_{3}$,  bandwidth ratio $r$  assuming $r> 1$ and temperature $T$ in a thermodynamic limit of a sufficiently long chain, where the continuum approach is applicable  (see discussion of discreteness in Sec. \ref{sec:Dictret}).  

At high temperatures $k_{\rm B}T>\hbar\Delta_{\rm tr}$ one can estimate the coupling matrix element for the processes relevant for the optical phonon overturn in Eq. (\ref{eq:SchrWollf}) as $\hbar V_{4}=\hbar V_{3}^2k_{\rm B}T/(r^3\Delta_{\rm opt}^2)$.  
The transition rate produced by the Fermi golden rule can be then estimated as 
\begin{eqnarray}
W_{4} \sim  \frac{V_{4}^2}{\Delta_{\rm tr}} =  C\frac{V_{3}^4(k_{\rm B}T/\hbar)^2}{r^7\Delta_{\rm opt}^{5}},  ~ C \approx 3\cdot 10^4, 
\label{eq:scattFGREst}
\end{eqnarray} 
where $C$ is the dimensionless proportionality constant determined numerically below in Sec.   \ref{sec:sumRellrgr}. 
Eq. (\ref{eq:scattFGREst}) is approximately  consistent with the numerical results for the phonon current relaxation rate  reported below in Sec. \ref{sec:num},   where the large absolute value of the   proportionality constant $C$ in Eq. (\ref{eq:scattFGREst})  is explained on a semiquantitative level.  The fourth order anharmonic interaction with longitudinal phonons should result in a relaxation rate behaving similar to Eq.  (\ref{eq:scattFGREst}) since longitudinal and transverse branches are weakly distinguishable for a large phonon wavevectors $k, q\sim \pi/a$ relevant for that regime.

\subsubsection{Backscattering at low temperature: multi-phonon emission }
\label{sec:subMultPh}

If the temperature is very low so there are no acoustic phonons  capable of  scattering initially excited optical phonon backwards, then the only remaining option for the optical phonon overturn is a  multi-phonon emission.  Such process is forbidden  for longitudinal acoustic phonons possessing the velocity exceeding that of optical phonons \cite{ab2023Cherenkov}, but it is allowed for transverse acoustic phonons even for $r\gg 1$ as shown below. 

As an illustration  consider the overturn of the optical phonon with the wavevector $k$ by means of the emission of $n$ transverse phonons with wavectors $q_{1}, q_{2},...q_{n}$ and total wavevector $\sum_{i=1}^{n}q_{i}=k$, which is the minimum wavevector change corresponding to the overturn that should take place simultaneously with the frequency change $\sum_{i=1}^{n}\omega_{\rm tr}(q_{i})= (\omega_{\rm opt}(q)-\omega_{\rm opt}(0))$  due to energy conservation law.  The most efficient emission takes place at the minimum number $n$  of emitted transverse phonons since the emission rate decreases exponentially as $\eta^{2n}$ with increasing $n$ (see Eq. (\ref{eq:AnhIntC}) for the definition of $\eta$).   
 
It can be shown rigorously for our specific phonon spectrum  Eq.  (\ref{eq:TrNN}) and generalized to any  monotonic function $\omega_{\rm tr}(q)$ that the optimum emission regime is realized for all identical emitted transverse phonons with wavectors $q_{i}=k/n$.  Assuming a large number of phonons $n \gg 1$, which is valid for $r \gg 1$ we can approximate the total emitted energy  as  $ \hbar\Delta_{\rm tr}a^2k^2/(4n)$ and  we estimate the minimum number of emitted phonons needed for the overturn as  
\begin{eqnarray}
n =   \eta_{\rm ph} r, ~~ \eta_{\rm ph}=\frac{k^2a^2}{4\sin(ka/2)^2}. 
\label{eq:NonlinNum}
\end{eqnarray}

Thus, in our model the number of transverse phonons needed to be emitted is proportional to the bandwidth ratio $r$ with the proportionality coefficient $\eta_{\rm ph}$ ranging from $1$ to $\pi^2/4\approx 2.5$.  The multiphonon emission rate $W_{\rm mult}$ is  proportional to the squared coupling constant for the $n$ phonon process. Consequently, we expect that 
\begin{eqnarray}
W_{\rm mult}  \propto \eta^{2\eta_{\rm ph}r}, 
\label{eq:MultRate}
\end{eqnarray}
where $\eta \sim 0.1$ Eq. (\ref{eq:AnhIntC}).  Additional dependence on the number of emitted phonons  can be originated from the wavevector dependence of the $n$-phonon interaction constant.   We do not investigate this dependence in detail, because  we were not able to reproduce this regime numerically as reported in Sec.  \ref{sec:quant}, so it cannot be compared to any experimental or numerical observation.

\subsection{Regimes of Fermi golden rule failure}
\label{sec:FGRBeyond}

Here we consider the regimes that cannot be characterized by  Fermi golden rule rates. These considerations are needed for understanding dynamics observed in numerical simulations reported below in Sec. \ref{sec:num} for very large or very small anharmonic interactions.   

The Fermi golden rule fails at very large anharmonic interactions $V_{3}$ in Eq. (\ref{eq:AnhIntC}), where the perturbation theory is no more applicable, and at very small anharmonic interactions,  where the discreteness of the phonon spectrum becomes important. Below we address both regimes. 

\subsubsection{Strong anharmonic coupling}
\label{sec:subStrongAnh} 

Consider the evolution of the initial harmonic state $a$ with certain wavevectors of  all  phonons.  The population $P_{a}(t)$ of this state  decreases with  time due to anharmonic interactions.  At short times it decreases as $P_{a}(t)=1-W_{*}^2t^2$, where $W_{*}=V_{*}/\hbar$ and  $V_{*}=\sqrt{\sum_{b\neq a}V_{ab}^2}$, where $V_{ab}$ is the off-diagonal matrix element of anharmonic interaction coupling harmonic states  $a$ with any other harmonic state $b$ possessing harmonic energies $E_{a}$ and $E_{b}$, respectively.     For a weak anharmonic interaction, where Fermi golden rule is applicable, this decay slows down at times much shorter than the effective minimum relaxation time $T_{*} \approx 1/W_{*}$,  since the squared time dependence for each specific state $b$ saturates at time $t \sim \hbar/|E_{a}-E_{b}|$.  For a strong interaction  the maximum rate $W_{*}$ gives a reasonable estimate for the relaxation rate consistent with the numerical results of  Sec. \ref{sec:num}. 

For the future comparisons with numerical results we report below   the  maximum relaxation rate  $W_{*}$ and its asymptotic behaviors at high and low temperatures evaluated for the anharmonic interaction Eq. (\ref{eq:AnhIntC})   as 
\begin{eqnarray}
W_{*}= \sqrt{\frac{1}{N}\sum_{q}4V_{3}^2\sin^2(qa/2)\left(1+\frac{2}{e^{\frac{\hbar\omega_{\rm tr}(q)}{k_{\rm B}T}}-1}\right)} =\begin{cases}
\sqrt{2}V_{3}, ~~ k_{\rm B}T \ll \Delta_{\rm tr}\\
4\sqrt{2}V_{3} \sqrt{\frac{k_{\rm B}T}{\Delta_{\rm tr}}}, ~~  k_{\rm B}T \gg \Delta_{\rm tr}.
\end{cases}
\label{eq:LargeVrel}
\end{eqnarray}

\subsubsection{Weak anharmonic coupling: effect of discreteness} 
\label{sec:Dictret}

The Fermi's golden rule is applicable to the continuous spectrum  in the infinite chain limit. The spectrum can be treated as continuous if the energy uncertainty associated with the specific state decay rate Eq. (\ref{eq:FGrSpec}) defined  as $ \hbar W_{\rm tot}$ exceeds the  interlevel splitting given by $\Delta E \approx (2\pi\hbar/L)$ max($|\partial\omega_{\rm tr}(k)/\partial k|$,  $|d\omega_{\rm opt}(k)/dk|$), where $L$ is the chain length.  Comparing two energies to each other we come up with the constraint on the strength of anharmonic interaction  $V_{3}$ needed for the Fermi's golden rule to be applicable, that can be expressed as 
\begin{eqnarray}
V_{3}> {\rm max}(1,  \sqrt{r}) \frac{\Delta_{\rm opt}}{\sqrt{N}}\begin{cases}
\sqrt{\frac{\pi(1+2r\cos(ka)+r^2)}{8}}, ~~ k_{\rm B}T \ll \hbar\Delta_{\rm tr}\\
\frac{\sin(ka)}{4}\sqrt{\frac{\pi r\hbar\Delta_{\rm opt}}{k_{\rm B}T}} ~~  k_{\rm B}T \gg \hbar\Delta_{\rm tr}.
\end{cases}
\label{eq:DiscrConstr}
\end{eqnarray}

What to expect if Eq. (\ref{eq:DiscrConstr}) is not satisfied and the  discreteness is significant? At  low temperature and reasonably short chains $N \sim 10$ the non-ergodic behavior  can emerge  \cite{LeitnerArnoldDiffusion97,LoganWolynes90,
Karmakar20VibrMBL,ab19FPU}
similarly to that found in  the earlier work \cite{ab2023Cherenkov}, where the interaction of optical phonons with longitudinal acoustic phonons was considered.  In the non-ergodic or many-body localized regime \cite{ab19FPU,Karmakar20VibrMBL} optical phonon current in a periodic chain will never relax to zero staying close to its initial value, except for the states with the total wavevector $0$ or $\pi/a$ where it disappears in average due to the inversion symmetry.  Yet in  the non-ergodic regime the current does not relax to zero but coherently oscillates switching between positive and negative values. 

There is a greater chance  to observe the localized regime for larger bandwidth ratio $r$ since the critical strength of anharmonic coupling separating discrete and quasi-continuous regimes grows with $r$ as $r^{3/2}$.  At high temperature the ergodic behavior should be restored due to phonon scattering. 
All these expectations are reasonably consistent with the results of numerical simulations reported in Sec. \ref{sec:num}.  



\section{Numerical Studies}
\label{sec:num}

The present section targets  verification of  the analytical expectations of Sec. \ref{sec:EnMomCons} using as accurate as possible numerical modeling of the optical phonon transport.  
We study how the  transport in periodic atomic chains   is  affected by interaction with transverse acoustic phonons using fully quantum mechanical or semi-classical approaches for periodic atomic chains.  The quantum mechanical treatment  reported  in Sec. \ref{sec:quant} is developed similarly to the earlier work \cite{ab2023Cherenkov} considering  longitudinal acoustic phonons.  It is limited to relatively short chains and small  total  energies (very low temperatures)  because of the exponential increase of the number  of significant quantum states with energy and number of sites $N$, which  prohibits an accurate numerical diagonalization for longer chains.  

The second approach (Sec.  \ref{sec:class}) treats acoustic phonons and their interaction with the optical phonon classically,  which permits us to characterize  large systems at high temperatures, where this consideration is reasonably justified.  


\subsection{Quantum mechanical treatment} 
\label{sec:quant}

The full quantum-mechanical treatment of optical phonon transport is quite similar to the earlier work \cite{ab2023Cherenkov} (see also Ref.  \cite{ab19FPU}).  Exact diagonalization of the system Hamiltonian Eq. (\ref{eq:FullH}) is not possible since the basis of its possible states  is infinitely large.  However,  for the states with relatively small energies we can limit our consideration to a number of acoustic phonons not exceeding a certain maximum number $n_{\rm max}$ and perform the full diagonalization with that limited basis \cite{ab19FPU,ab2023Cherenkov}.   Our approximation can be validated considering the dependence of  certain  parameters of interest on the number $n_{\rm max}$.  In most of the calculations we investigated the time evolution of the probability  $P_{n}(t)$ to find the excited optical phonon in its initial state with a certain  wavevector $k=2\pi n/L$ and we used its infinite time limit $P_{n}(\infty)$  as a convergence control parameter.  

The calculations reported below are performed for a periodic chain of $N=12$ sites  and the anharmonic interaction strength $V_{3}\leq 0.5\Delta_{\rm opt}$. In this specific case a good convergence for probabilities $P_{n}(\infty)$  is obtained for $n_{\rm max}=6$, which corresponds to the basis of $12376$ states.  The difference of the results for $n_{\rm max}=5$ and $n_{\rm max}=6$ is always less than few percents and their difference for  $n_{\rm max}=6$ and $n_{\rm max}=7$ is less than $1$\%. Therefore, we believe that our method gives a good approach to the  actual quantum  evolution of the system.  

We investigate the relaxation of the current for the initially excited optical phonon with the wavevector $k=2\pi n/(aN)$.  The state with  $n=N/4=3$ is  chosen because it is located in the middle of the band  and possesses a maximum phonon velocity $v(k)=\Delta_{\rm opt}\sin(ka)/2$ realized at $ka=\pi/2$ Eq. (\ref{eq:EnFlux}).  The results for other initial wavevectors are quite similar, except for $n=0$ where non-ergodic behavior was seen for $V < 0.5\Delta_{\rm opt}$ for any considered bandwidth ratios because it is the lowest energy optical phonon state having no decay channels.   Since the observed dependence on the initial wavevector is quite similar to that reported in Ref. \cite{ab2023Cherenkov} we do not show the results of calculations for $n\neq 3$.

We investigated two different initial states including (A) the initial state with a single excited optical phonon with the wavevector $k=2\pi n/L$ and  (B) the two-phonon  initial state with a single excited optical phonon with the  same wavevector $k$ and a single acoustic phonon with the wavevector $-k$. The case  (A) earlier used in Ref. \cite{ab2023Cherenkov} is targeted to examine the relaxation due to the emission of acoustic phonons, while in the case (B) we also investigate the effect of resonant  backscattering on the optical phonon current for a large bandwidth ratio $r>1$.   The    more complicated initial conditions are briefly discussed in the end of the present section.

\subsubsection{Single excited optical phonon} 
\label{sec:subSingleIn}

\begin{figure}%
    \centering
    \subfloat[\centering ]{{\includegraphics[width=7.5cm]{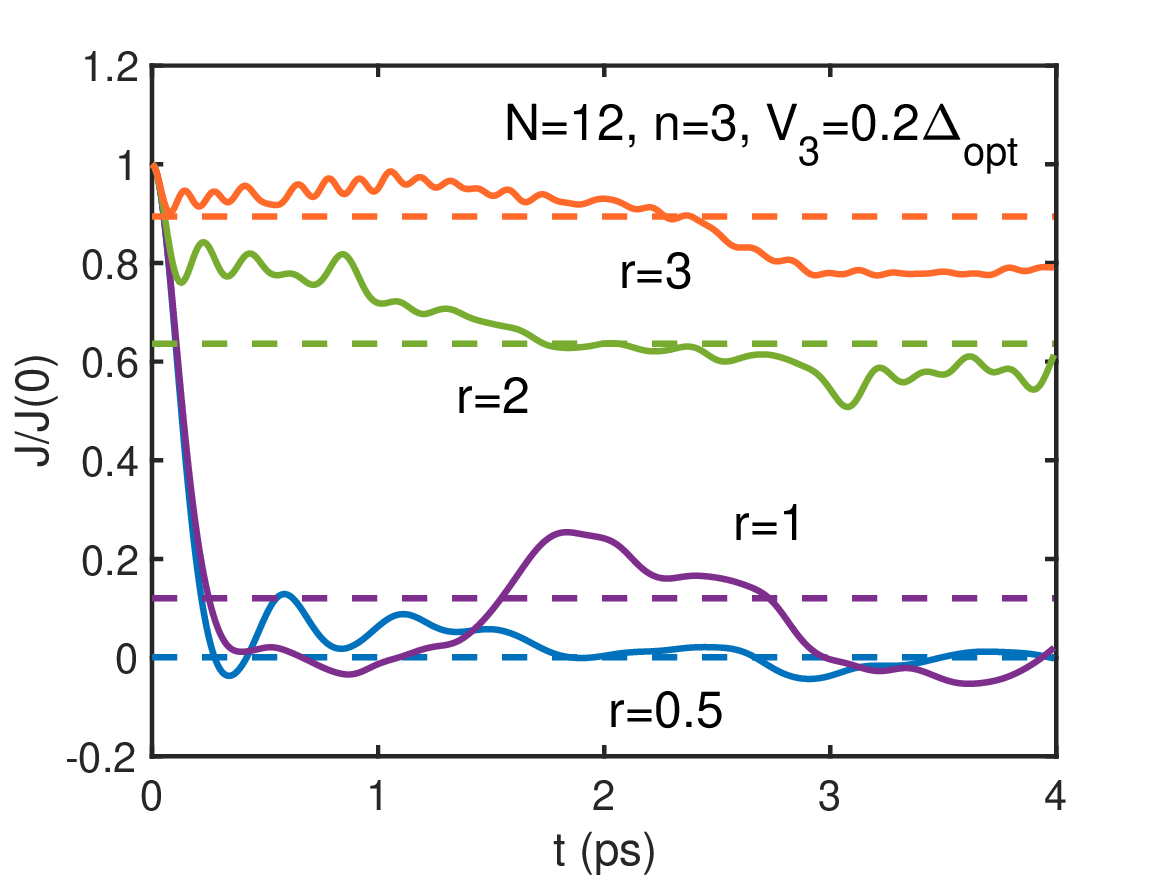} }}%
    \qquad
    \subfloat[\centering ]{{\includegraphics[width=7.5cm]{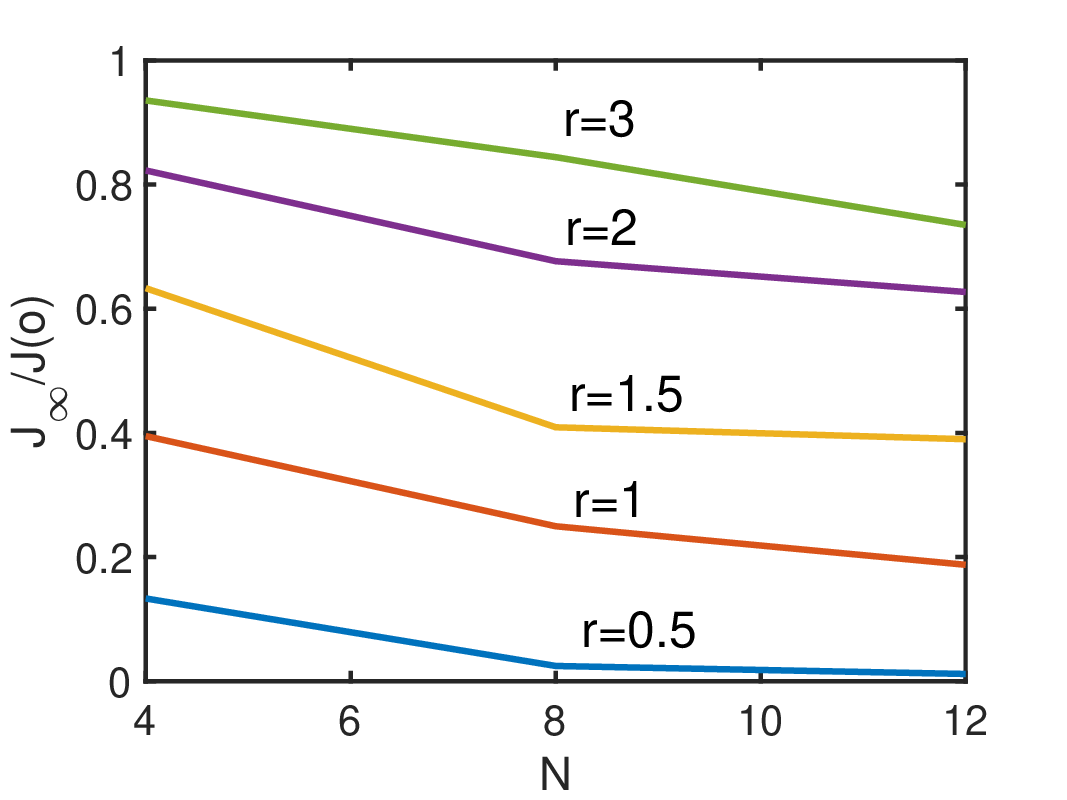} }}%
    \caption{Time dependence  of currents (a) and size dependence of time averaged currents $J_{\infty}$  (b) for different ratios $r$ of transverse acoustic and optical bandwidths.  }%
    \label{fig:RelaxrtDep}%
\end{figure}

The time evolution of the optical phonon current for the initial state  containing only one excited optical phonon with the wavevector $k=\pi/(2a)$ ($N=12$) is shown in Fig. \ref{fig:RelaxrtDep}.a  for four different  bandwidth ratios $r$.  We choose the maximum coupling strength $V_{3}=0.2\Delta_{\rm opt}$, where the perturbation theory with respect to anharmonic interaction is still applicable.    
Infinite time limits of current shown by dashed lines are  evaluated averaging current over time, that sets all oscillating contributions to zero as in Refs.  \cite{ab2023Cherenkov,BerezinskiiGorkov79}. 

The observed evolution of currents   for bandwidth $r\leq 1$ and $r>1$ are clearly different.  For  $r\leq 1$ the current rapidly decreases to the small value compared to the  initial current $I(0)$,  while for $r=2$ and $3$ it changes weakly compared to its initial value.  We interpret this difference as the separation  between ergodic ($r=0.5$ and $1$) and non-ergodic ($r=1.5$, $2$ and $3$) regimes. In an ergodic regime  the time averaged current represented by its infinite time limit  should approach its thermodynamic average of $0$ in the thermodynamic limit of an infinite system $N \rightarrow \infty$.  

We evaluated the size ($N$) dependence of a time-averaged current $J_{\infty}$ corresponding to its infinite time limit as reported in Fig. \ref{fig:RelaxrtDep}.b.  The size dependence, indeed, shows fast reduction of $J_{\infty}$ with the system size for $r\leq 1$ and a substantially weaker size dependence for $r >1$ that is consistent with our assumption of the transition between ergodic and non-ergodic behaviors at $r \approx 1$.    However, we expect that at larger sizes $N$ ergodic behavior should be restored for an arbitrary $r$ due to forward scattering and higher order anharmonic interactions. This contrasts to the case of optical phonon interaction with longitudinal phonons where any transitions for a single excited optical phonon are forbidden in the regime of Cherenkov's constraint \cite{ab2023Cherenkov}. 

\begin{figure}%
    \centering
    \subfloat[\centering ]{{\includegraphics[width=7.5cm]{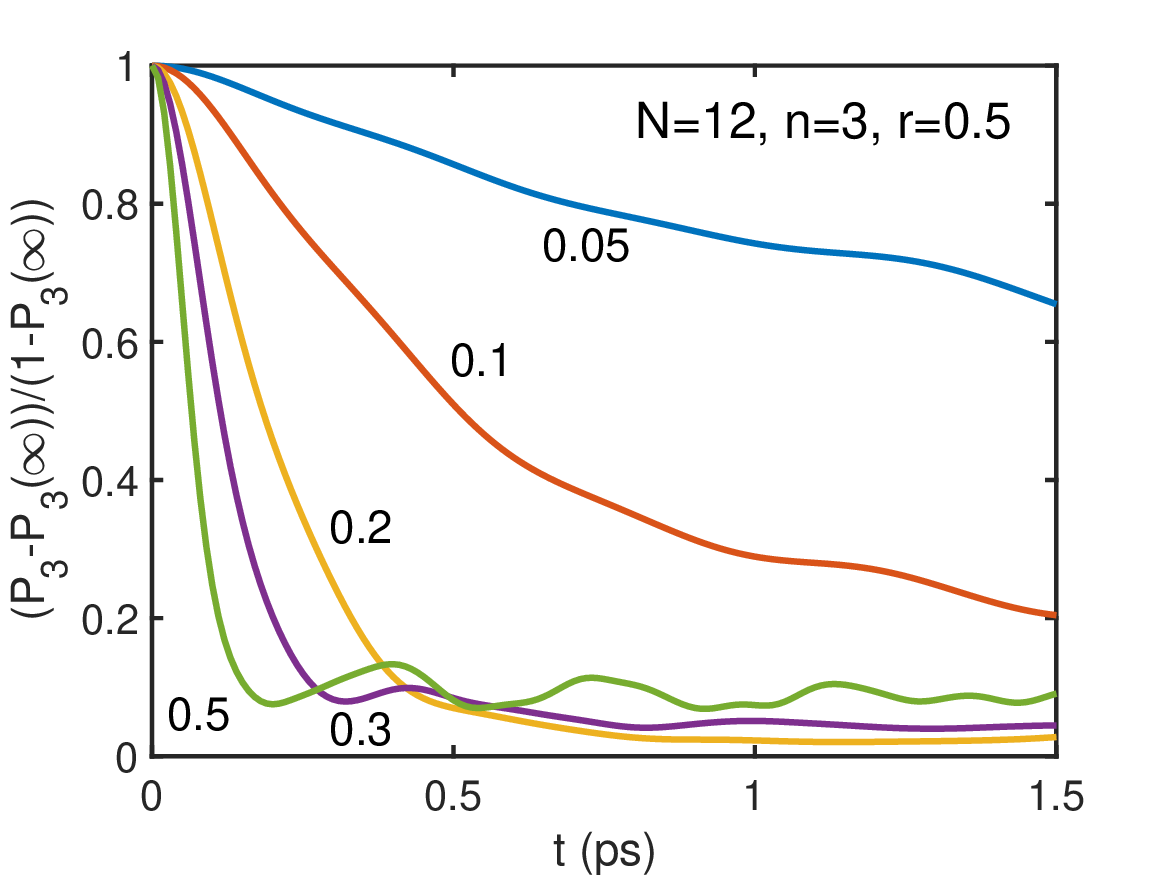} }}%
    \qquad
    \subfloat[\centering ]{{\includegraphics[width=7.5cm]{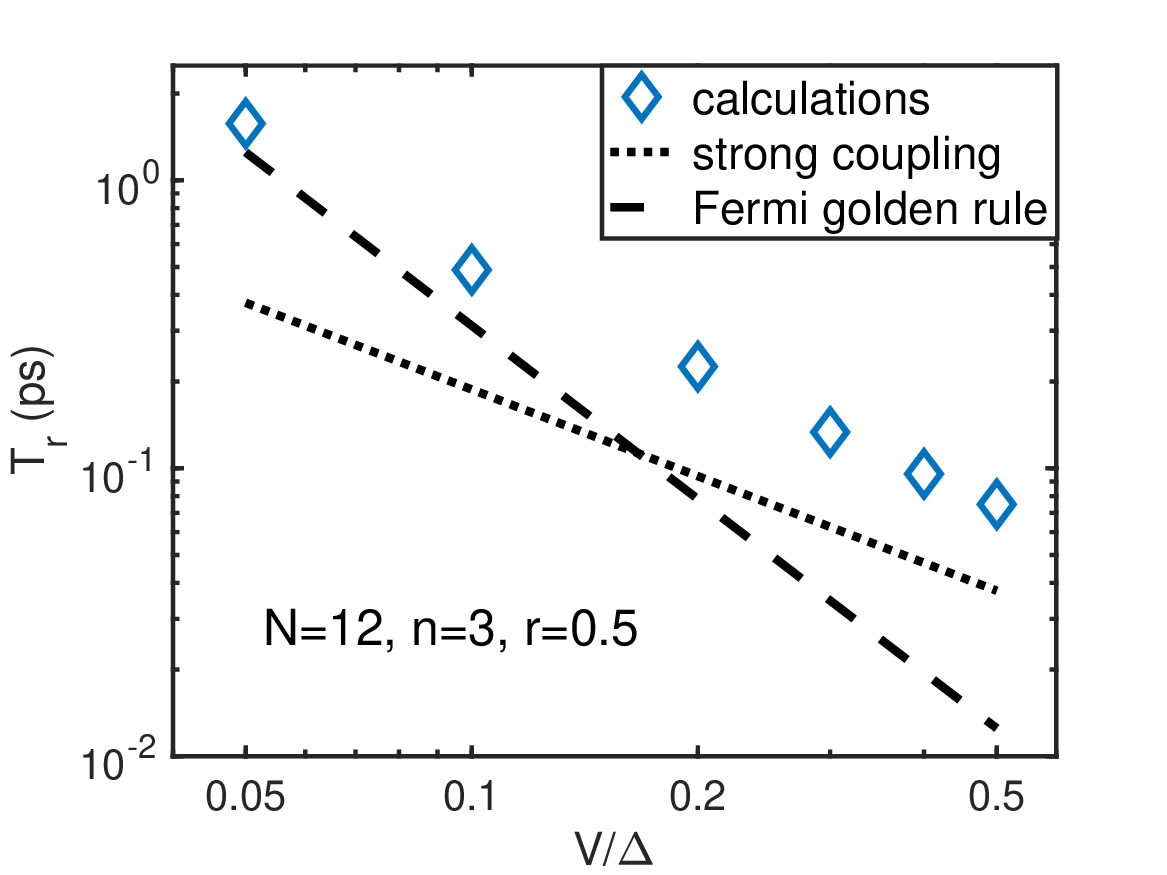} }}%
    \caption{(a) Relaxation of the population of the optical phonon midband state $n=3$ for $12$ site chain, and bandwidth ratio $r=0.5$ at different strengths of anharmonic interaction. (b)  Dependence of population relaxation time on the strength of anharmonic interaction shown together with the rate predicted by the Fermi's golden rule (\ref{eq:FGrSpec}) shown by the dashed line and characteristic minimum lifetime $1/W_{*}$ Eq. (\ref{eq:LargeVrel}) shown by the dotted line.}%
    \label{fig:CurClLta}%
\end{figure}

We  investigated the relaxation of the population of the initially excited state $P_{n}(t)$  in the ergodic regime $r=1/2$ to compare relaxation times estimated using Fermi's golden rule  (inverse relaxation rate  $W_{\rm tot}$ Eq. (\ref{eq:FGrSpec})) with their numerical estimates.   The numerical relaxation time $T_{\rm rel}$ is defined using time dependent populations $P(t)$  shown in Fig.  \ref{fig:CurClLta}.a setting $P(T_{\rm rel})=P_{\infty}+(1-P_{\infty})e^{-1}$ for $n=3$.  Numerical relaxation times are shown by diamonds in Fig.  \ref{fig:CurClLta}.b vs.  anharmonic coupling strengths and compared to the Fermi's golden rule estimate $W_{\rm tot}^{-1}$ (Eq. (\ref{eq:FGrSpec}), dashed line) and the minimum relaxation time  $W_{*}^{-1}$ realized for a strong anharmonic coupling  (Eq. (\ref{eq:LargeVrel}), dotted line).  It turns out that relaxation time switches between two regimes with increasing the coupling strength $V_{3}$ as expected.  For $V_{3}<0.03\Delta_{\rm opt}$  the relaxation is no more seen, but  instead the current oscillates around its average value close to its initial value. This suggests an ergodicity breakdown emerging at $V_{3}<0.03\Delta_{\rm opt}$  with the current evolution similar to observed behaviors for $r\geq 2$ as shown in Fig. \ref{fig:RelaxrtDep}. a.    

Non-ergodic behaviors  are caused by  discreteness as discussed in Sec. \ref{sec:Dictret}.  An increase of the number of sites $N$ should most probably eliminate that behavior even for $r>1$  in a thermodynamic limit of $N\rightarrow \infty$ due to allowed forward scattering that was  lacking in the previous study  \cite{ab2023Cherenkov}. 



\subsubsection{Two phonon initial state} 
\label{sec:sub2Ph} 

To investigate the effect of a resonant two-phonon interaction onto the system dynamics we modified the initial conditions compared to Sec. \ref{sec:subSingleIn} using  an optical phonon with the wavevector $k=2\pi n/L$ ($n=3$, $N=12$) and one transverse acoustic phonon with the opposite wavevector $-k$.  
We focus  only on a nonergodic regime $r>1$ since in the ergodic regime the current evolution is practically the same as for the initial conditions of a single excited optical phonon reported  in Sec.  \ref{sec:subSingleIn}. 

For two-phonon  initial condition the time-averaged optical phonon current approaches zero because of the  inversion symmetry of the problem for a total wavevector equal $0$ suggesting that each eigenstate of the problem is composed by the symmetric or antisymmetric combination of pairs of states with all opposite wavevector like $|k,-k>$ and $|-k,k>$ for two-phonon state.  The initial state used in Sec. \ref{sec:subSingleIn} possesses a wavevector $k=\pi/(2a)$. The conservation of the wavevector does not allow a system transition to the symmetric  state with the opposite wavevector, since the difference of two state  wavevectors $\pi/a$ is not equal to an integer number of inverse lattice periods $2\pi/a$.   Consequently,   
the inversion symmetry is broken and the average current differs from zero.       

\begin{figure}%
    \centering
    \subfloat[\centering ]{{\includegraphics[width=7.5cm]{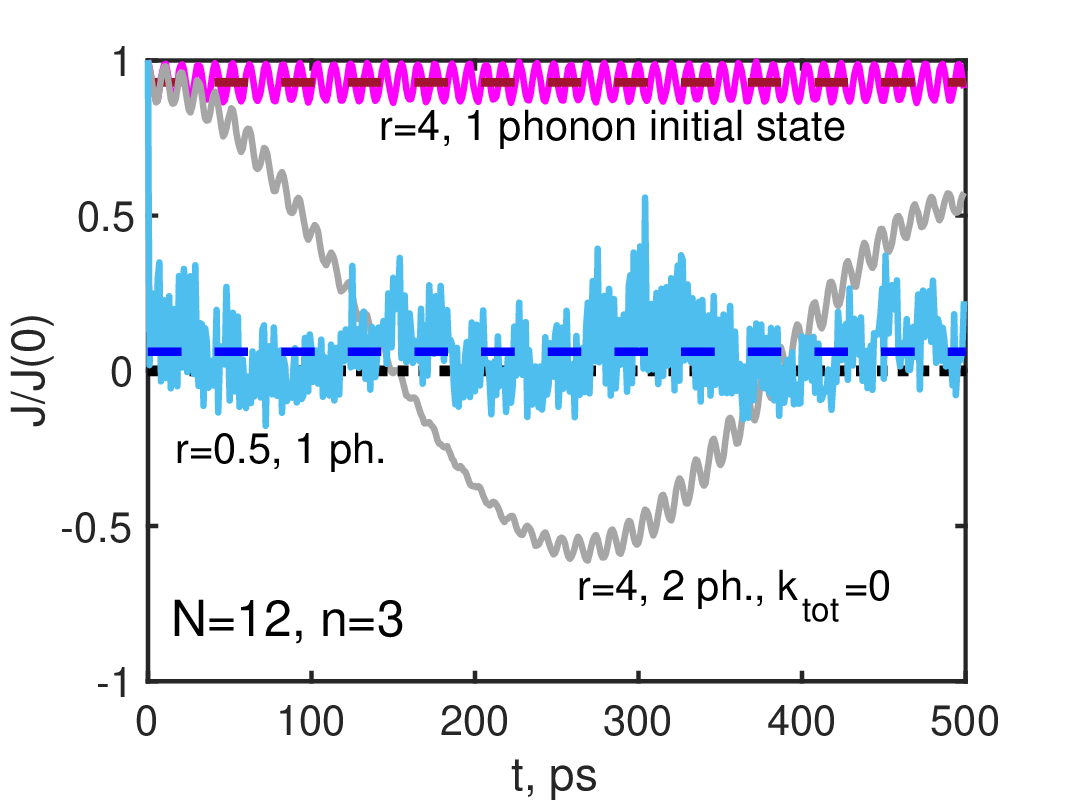} }}%
    \qquad
    \subfloat[\centering ]{{\includegraphics[width=7.5cm]{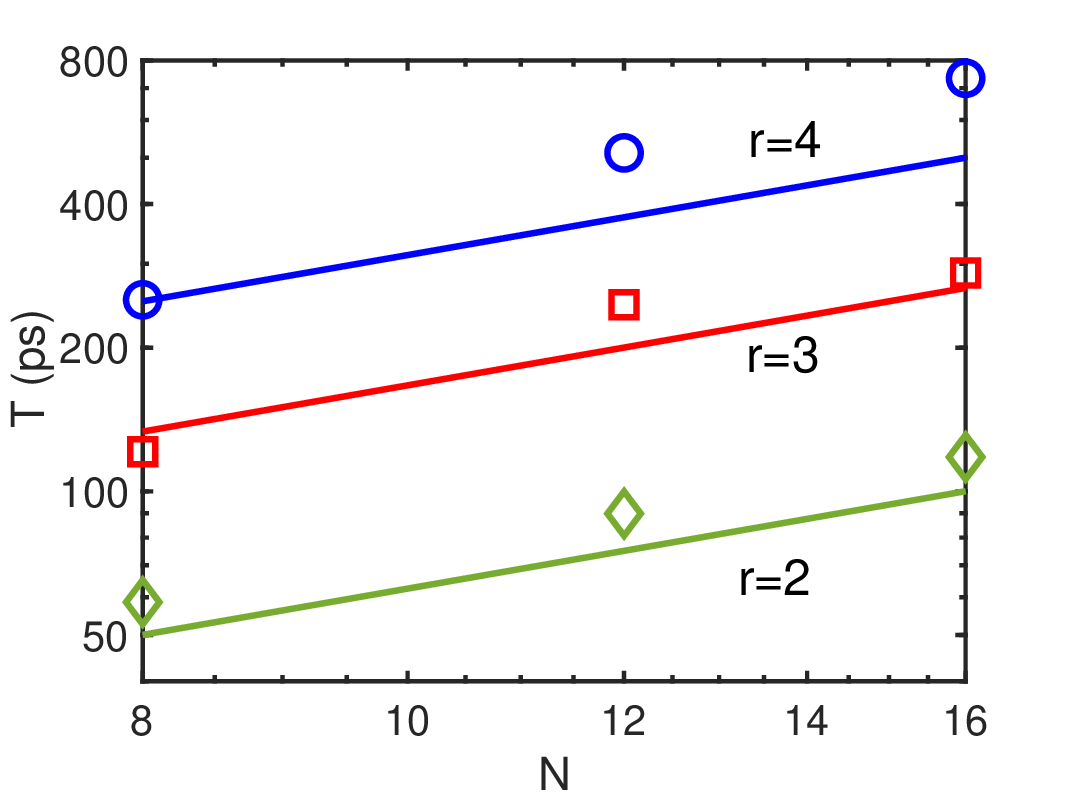} }}%
    \caption{Coherent oscillation of current for a two phonon initial state with the total wavevector $0$.  (a)  Dependence of current on time. (b) Dependence of oscillation period on chain length at different  ratios of bandwidths $r$. Straight lines show the expected oscillation period due to  resonant interactions in Eq. (\ref{eq:Period}).  Anharmonic coupling strength is $V_{3}=0.2\Delta_{\rm opt}$.}%
    \label{fig:CurClLt2}%
\end{figure}

For two phonon initial condition and $r\geq 2$ the current shows nearly coherent  oscillations around zero,  as illustrated  in Fig. \ref{fig:CurClLt2}.a (grey line) for a bandwidth ratio $r=4$, in a stark contrast with the time evolution  for a single phonon initial condition,  as shown in the same graph (magenta line).   Similar oscillations were found for a wide range of anharmonic coupling strengths and bandwidth ratios $r\geq 2$.

 To interpret the observed coherent  oscillations we employ the secular perturbation theory limiting our consideration to the  only two resonant states $|k,-k>$ (the initial state) and $|-k, k>$ (the symmetric state) possessing identical harmonic energies and coupled by the fourth order anharmonic interaction $V(k, -k, -k, k)=16\hbar V_{3}^2/(N\Delta_{\rm opt}(r^2-1))$  Eq. (\ref{eq:SchrWlfRes}).  For the given initial  condition the system coherently oscillates between these two states possessing opposite optical phonon currents with the period 
\begin{eqnarray}
T_{\rm res}=\frac{2\pi \hbar}{V(k,-k,-k,k)}=\frac{2\pi (r^2-1)N\Delta_{\rm opt}}{16V_{3}^2}. 
\label{eq:Period}
\end{eqnarray}

In Fig. \ref{fig:CurClLt2}.b we show  the dependence of  numerically estimated oscillation periods (symbols) on a system size and bandwidth ratios.  Here we used the maximum number of acoustic phonons equal to $5$ which is justified for $N\leq 12$ but questionable for $N=16$.  Numerical estimates  are consistent  with the  theory  predictions Eq. (\ref{eq:Period}) shown by the straight lines.    Coherent oscillations of current suggest a non-ergodic behavior in spite of average current vanishing.  

We also examined more complicated initial conditions. If one use three phonons including one optical phonon, one acoustic phonon with an opposite wavevector  and one additional acoustic phonon, the time averaged current approaches an intermediate value between $0$ and its initial value $J(0)$ (for instance, if we add one more transverse acoustic phonon with the same wavevector  $-k$ we got $J_{\infty}\approx J(0)/2$) in the limit of small anharmonic coupling. This is a typical behavior for a non-ergodic regime.  In the ergodic regime realized for long chains $N\gg 1$ and/or high temperatures $k_{\rm B}T \gg \hbar\Delta_{\rm opt}$ the fourth order anharmonic  interaction Eq.  (\ref{eq:SchrWlfRes})  leads  to the  relaxation investigated  below in Sec.  \ref{sec:class} using the semiclassical approach. 

 \subsection{Semiclassical treatment} 
\label{sec:class}

\subsubsection{Semiclassical Model}
\label{sec:class-mod}

Since most of experiments are performed at room temperature the assumption of a low temperature made in the previous section is  not satisfied there.  At room temperature a typical thermal energy $k_{\rm B}T$ is comparable to the acoustic phonon bandwidth and exceeds the optical phonon bandwidth. Under those conditions 
\begin{eqnarray}
\Delta_{\rm opt}, ~ \Delta_{\rm tr} < k_{\rm B}T
\label{eq:HighT}
\end{eqnarray}
we employ a semiclassical treatment of acoustic phonons  to simplify the consideration.   Here we use it in the form of  Refs. \cite{Senthilkumar2005DNASemicl} developed to investigate charge transfer in DNA.  

The semiclassical  approach can be formulated as follows. We describe the optical phonon by the wavefunction within the momentum (wavevector) representation 
\begin{eqnarray}
|\psi> = \sum_{k}c_{k}|k>, 
\label{eq:semclwf}
\end{eqnarray}
where symbols $|k>$ stands for the state with the given wavevector $k$ and coefficients $c_k$ are wavefunction amplitudes determining probabilities $P_{k}=|c_{k}|^2$ to find the phonon in the state with the given wavevector $k$, which are subject to the normalization condition $\sum_{k}P_{k}=1$.  Transverse phonons are characterized by classical coordinates $u_{q}$ and momenta $p_{q}$ also defined in the wavevector representation.  

The system is described by the classical Hamiltonian  Eq. (\ref{eq:FullH}) 
\begin{eqnarray}
H=\sum_{q}\frac{p_{q}p_{-q}}{2M}+\frac{M}{2}\sum_{q}\omega_{\rm tr}(q)^2u_{q}u_{-q}+\sum_{k}\hbar\omega_{\rm opt}(k)c_{k}^{*}c_{k} +
\nonumber\\
+2V_{3}\sqrt{2\hbar M\Delta_{\rm tr}}\sum_{k,q}\sin(qa/2)^2u_{q}(c_{k}^{*}c_{k-q}+C.C.). 
\label{eq:HamCl}
\end{eqnarray}
The time evolution of the variables is defined by the Hamiltonian  equations of motion 
\begin{eqnarray}
\frac{dc_{k}}{dt}=-\frac{i}{\hbar}\frac{\partial H}{\partial c_{k}^{*}}, ~ \frac{dc_{k}^{*}}{dt}=\frac{i}{\hbar}\frac{\partial H}{\partial c_{k}},~
\frac{du_{k}}{dt}=-\frac{\partial H}{\partial p_{-k}}, ~ \frac{dp_{k}}{dt}=\frac{\partial H}{\partial u_{k}}.
\label{eq:EqMotCl}
\end{eqnarray}
It is straightforward to check that these equations conserve both the total energy $H$ and the optical phonon wavefunction normalization $\sum_{q}|c_{k}|^2=1$.  This approximation  treats classically all participating phonons. 

A semiclassical approach is justified for  optical phonons if  anharmonic interactions are weak compared to harmonic ones 
\begin{eqnarray}
V_{3}\ll \Delta_{\rm opt}, \Delta_{\rm tr}. 
\label{eq:EqClassic2}
\end{eqnarray}
In this regime one can think of an optical phonon as a plane wave with a certain wavector $k$ undergoing rare scattering events.  This  justifies averaging of  the nonlinear  interactions over the optical phonon wavefunction   implied  in the classical equations Eq. (\ref{eq:EqMotCl}).  The semiclassical approach can miss the effect of quantum discreteness \cite{LeitnerArnoldDiffusion97}, though the analysis of the minimalist system of weakly anharmonically coupled oscillators shows similarity between classical and quantum considerations \cite{Cohen13}. 


Below we focus on modelling the relaxation in the regimes of the smallest numbers of sites $N$,  where Fermi's golden rule in the form of Eq. (\ref{eq:FGrSpechT})  ($N=20$ and $r < 1$) or Eq. (\ref{eq:scattFGREst}) ($N=48$ and $r >1$)  is still applicable for a reasonable anharmonic coupling strength $V_{3} \geq 0.01 \Delta_{\rm opt}$.  In both cases we choose the initial midband optical phonon state with the wavevector $k=2\pi n/L$ and $n=N/4$ with one exception related to the investigation of  the dependence of the relaxation rate on the initial state. 
The dynamics  for smaller numbers of sites   is slower and contains oscillations   determined by  rare resonances similarly to Ref.   \cite{Lvov15}.  Its consideration is beyond the scope of the present work.  

In our calculations we set the thermal energy equal to the double bandwidth for transverse acoustic phonons $k_{\rm B}T=2\hbar\Delta_{\rm tr}$.  At that temperature the classical approximation is reasonably justified, while the temperature under consideration is not much higher than the room temperature where $k_{\rm B}T\sim \hbar\Delta_{\rm tr}$. Consequently,  our results are valid  at room temperature  at least qualitatively. 

\subsubsection{Results of calculations}
\label{sec:class-res}

At time $t=0$ the optical phonon is  placed to the state with a certain wavevector $k$ similarly to Sec.  \ref{sec:quant}.   For transverse acoustic phonons we choose zero initial coordinates $u_{q}=0$ and random momenta $p_{q}=p_{-q}=\xi_{q}\sqrt{2Mk_{\rm B}T}$, where $\xi_{q}$ are random numbers generated from the Gaussian distribution with a unit width. These initial conditions introduce the relevant temperature  if anharmonic interaction is weak Eq. (\ref{eq:EqMotCl})  so kinetic and potential  energies for each normal  mode should be approximately equal to  $k_{\rm B}T/2$.  

Time dependencies of probabilities $P_{k}(t)=|c_{k}(t)|^2 $ are calculated solving the Hamiltonian equations Eq. (\ref{eq:EqMotCl}) and  averaging their solutions  over many random initial conditions corresponding to the given temperature until the accuracy of few percents is obtained.  Below the results of calculations are reported for  time dependent probabilities $P_{k}$ for different system sizes, anharmonic coupling strengths  and bandwidth ratios $r$.  

First, we report time dependent currents for representative chain of $N=20$ sites with initially excited midband optical phonon.  The anharmonic coupling was chosen as $V_{3}=0.05\sqrt{\Delta_{\rm opt}\Delta_{\rm tr}}$ to have  bandwidth independent   Fermi's golden rule rates of emission and absorption  Eq. (\ref{eq:FGrSpechT}).    The absolute value of $V_{3}$ and the system size were chosen to avoid discreteness and have perturbation theory applicable.  Remember that the temperature is  given by  $k_{\rm B}T=2\hbar\Delta_{\rm tr}$. 

\begin{figure}
\includegraphics[scale=0.5]{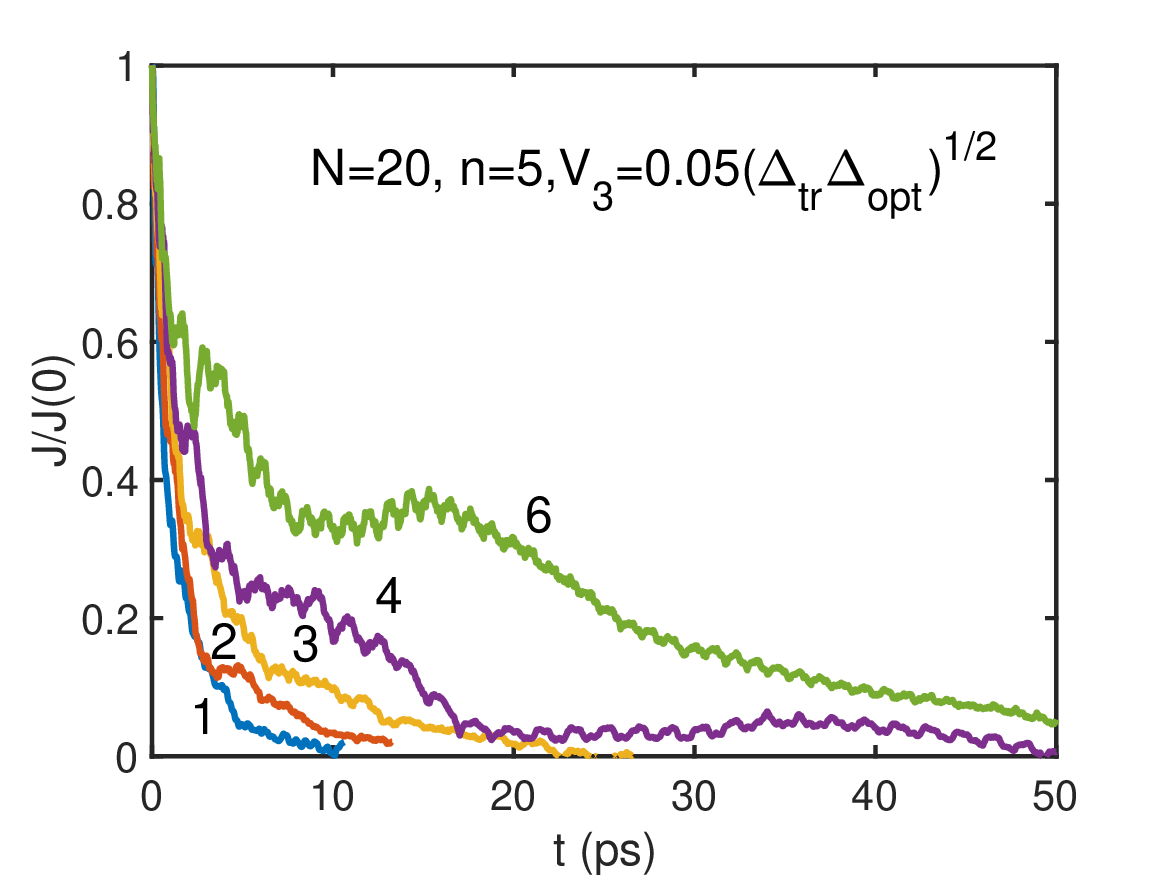} 
\caption{Time dependence of current for different ratios ($r$) of acoustic to optical bandwidth indicated near each graph.}. 
\label{fig:CurClLt}
\end{figure}

Time dependencies of current are very different for $r\leq 1$ where it  can be characterized by a single relaxation time  and $r\geq 2$ where it relaxes   in two stages including the fast initial relaxation during the time of the order of one picosecond and a slow subsequent relaxation taking  orders of magnitude longer.   This observation is consistent with the expectations of Sec.  \ref{sec:FGRRates}.    For $N=20$ and $r>1$ there are some discreteness effects that are seen if Fig. \ref{fig:CurClLt} as damped oscillations emerging due to three or four phonon resonances. We do not consider them in detail.   

To characterize both  regimes we examine relaxation time behaviors separately in Sec.  \ref{sec:sumRelsmr} for a small bandwidth ratio $r=1/2$ and in Sec. \ref{sec:sumRellrgr} in the opposite regime of $r>1$. 

\subsubsection{Relaxation at small bandwidth ratio ($r=1/2$)}
\label{sec:sumRelsmr}

\begin{figure}%
    \centering
    \subfloat[\centering ]{{\includegraphics[width=7.5cm]{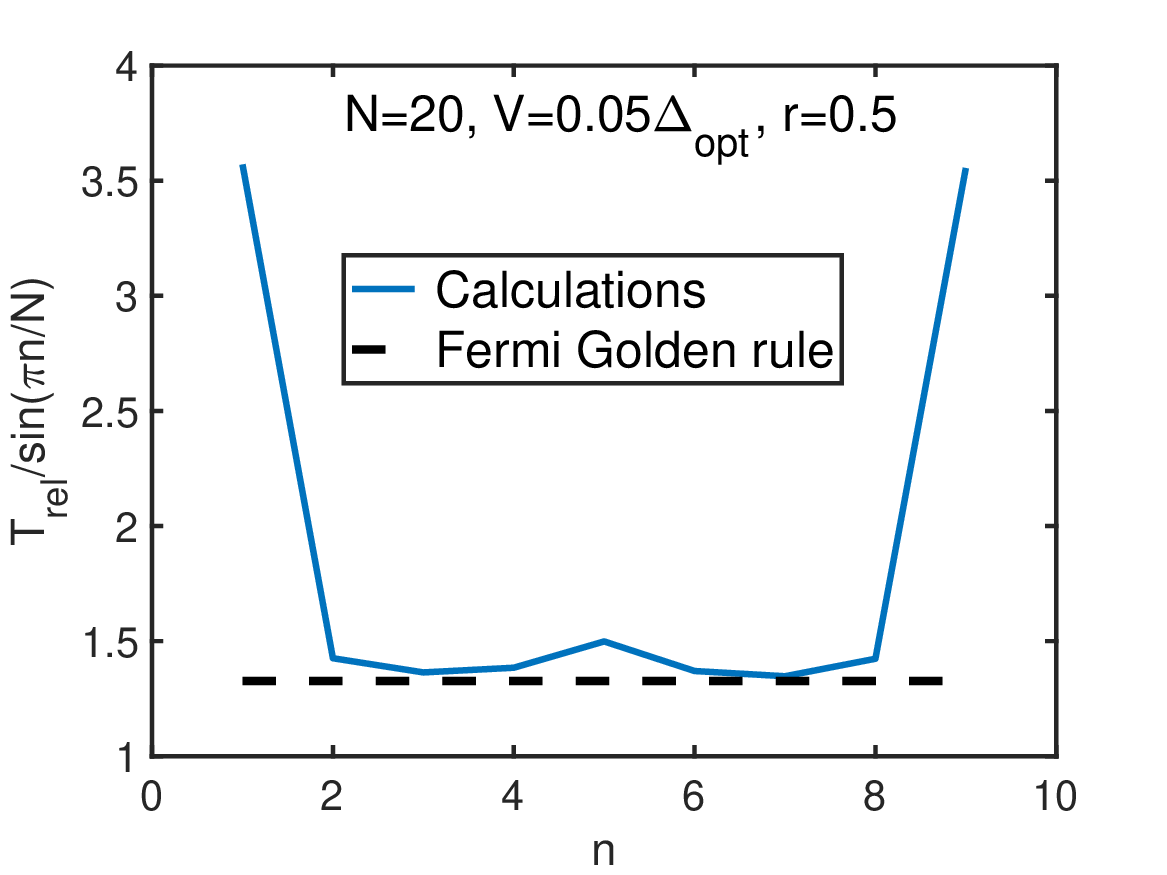} }}%
    \qquad
    \subfloat[\centering ]{{\includegraphics[width=7.5cm]{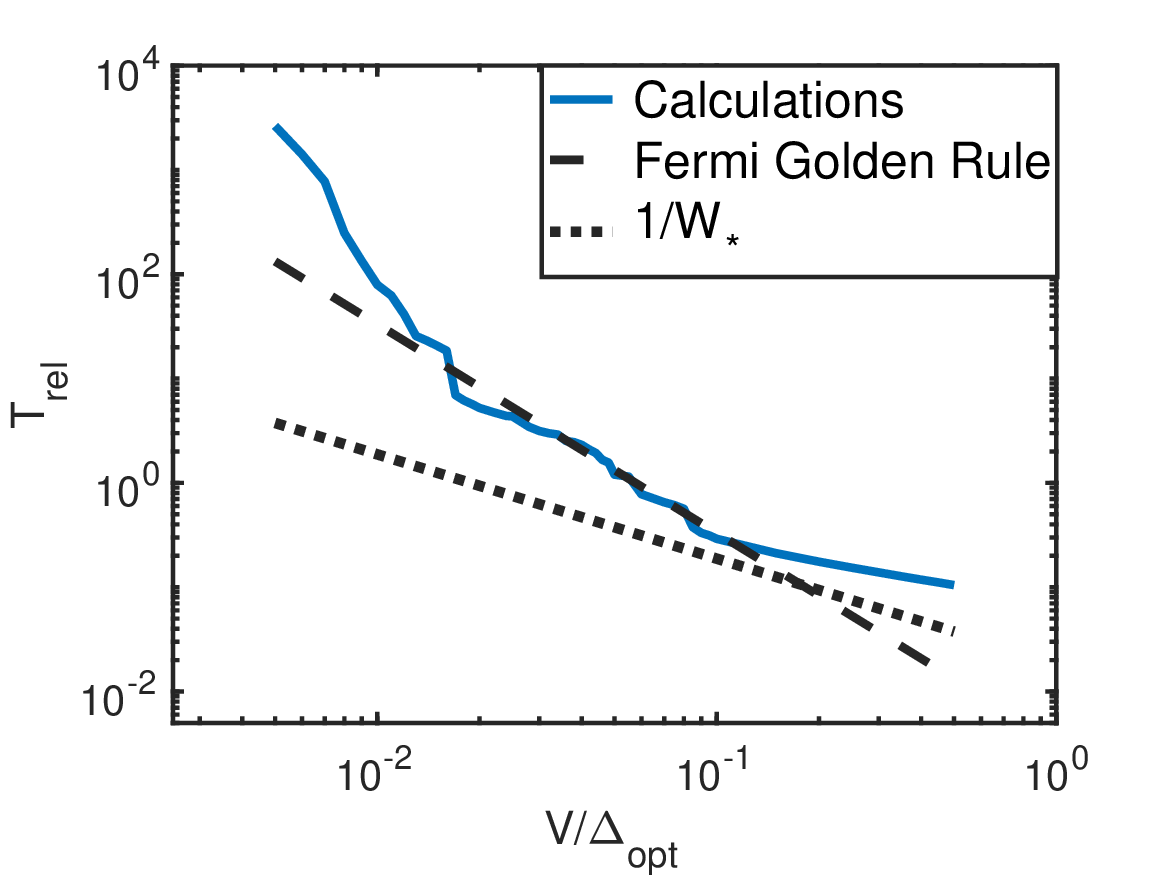} }}%
    \caption{(a) Dependence of the rescaled relaxation time on the wavevector  for $N=20$  compared to the Fermi's golden rule result  Eq.  (\ref{eq:FGrSpechT}) for the parameters $k_BT=2\Delta_{\rm tr}$, $V=0.05\Delta_{\rm opt}$,  $r=0.5$.  (b) Dependence of the relaxation time on the strength of the anharmonic interaction for the same parameters as in (a) and $n=5$.  Results are compared with  the Fermi's golden rule Eq. (\ref{eq:FGrSpechT}) (dashed line)  and the minimum relaxation time $1/W_{*}$ Eq. (\ref{eq:LargeVrel}) (dotted line).}%
    \label{fig:FGRCls}%
\end{figure}

For $r\leq 1$ it is natural to expect a population relaxation following the Fermi's golden rule predictions at high temperature Eq. (\ref{eq:FGrSpechT}).  To verify that, we examined the population relaxation for the bandwidth  ratio $r=1/2$ and all possible initial wavevectors $k$ for $N=20$  as reported in Fig.  \ref{fig:FGRCls}.  The relaxation time for each initial state time dependent population $P_{n}(t)$ corresponding to the wavevector $k=2\pi n/L$ was estimated setting $P_{n}(T_{\rm rel}(n))=P_{n}(\infty)+(1-P_{n}(\infty))e^{-1}$, where $P_{n}(\infty)\approx 1/N$ is the infinite time limit of the phonon state population in the classical ergodic system.  Exact time averaged  populations slightly differ from each other for different wavevectors, but this difference is negligible for the temperature under consideration due to the conservation of the number of optical phonons and their classical treatment.  

To compare the numerical results for relaxation times with the theory Eq. (\ref{eq:FGrSpechT})  predicting that 
$$
T_{\rm rel}(n)=\frac{\hbar\Delta_{\rm tr}\Delta_{\rm opt}\sin(|n|\pi/N)}{16V_{3}^2k_{\rm B}T}
$$
we plot the ratios $T_{\rm rel}(n)/\sin(n\pi/N)$ vs $n$ which should be independent of the initial wavevector in Fig.  \ref{fig:FGRCls}.a.    Only half of initial wavevectors is shown since the data for $n$ and $-n$ are identical due to the inversion symmetry.  We did not examine lifetimes of the states with $n=0$ and $N/2$ where the Fermi's golden rule must obviously fail.  

According to our observations the results for the relaxation rate are in an excellent agreement with the theory for $N=20$ and $2\leq n \leq 8$.
The deviations at small wavevectors are probably because the predicted relaxation rate is too fast there, so the Fermi's golden rule is no longer  applicable   (see discussion in Sec.  \ref{sec:subStrongAnh}). 
The dependence of the relaxation time on the anharmonic interaction  $T_{\rm rel} \propto V_{3}^{-2}$ is also consistent with the theory predictions except for very small coupling constants $V_{3} \leq 0.01\Delta_{\rm opt}$ where discreteness becomes significant as shown in Fig. \ref{fig:FGRCls}.b. 

\subsubsection{Relaxation at large bandwidth ratio $r>1$}
\label{sec:sumRellrgr}

 For a large bandwidth ratio $r \geq 2$ the evolution of current cannot be described using a single relaxation time as it is clearly seen in Fig.  \ref{fig:CurClLt}.   The current relaxes in two stages including the first fast stage taking few picoseconds due to   absorption or emission  Eq. (\ref{eq:FGrSpechT}) (forward scattering) and the second slow stage that is expected   to be determined by much slower phonon scattering Eq. (\ref{eq:scattFGREst}) (backscattering).  Here we report the results for the slow stage relaxation for  $N=48$ and  compare the current evolution at the slow stage  with the predictions of theory Eq. (\ref{eq:scattFGREst}).  We always excite initially the midband optical phonon $n=12$ and use the temperature corresponding to the thermal energy exceeding the transverse phonon bandwidth twice $k_{\rm B}T=2\hbar\Delta_{\rm tr}$. 
 
\begin{figure}%
    \centering
    \subfloat[\centering ]{{\includegraphics[width=7.5cm]{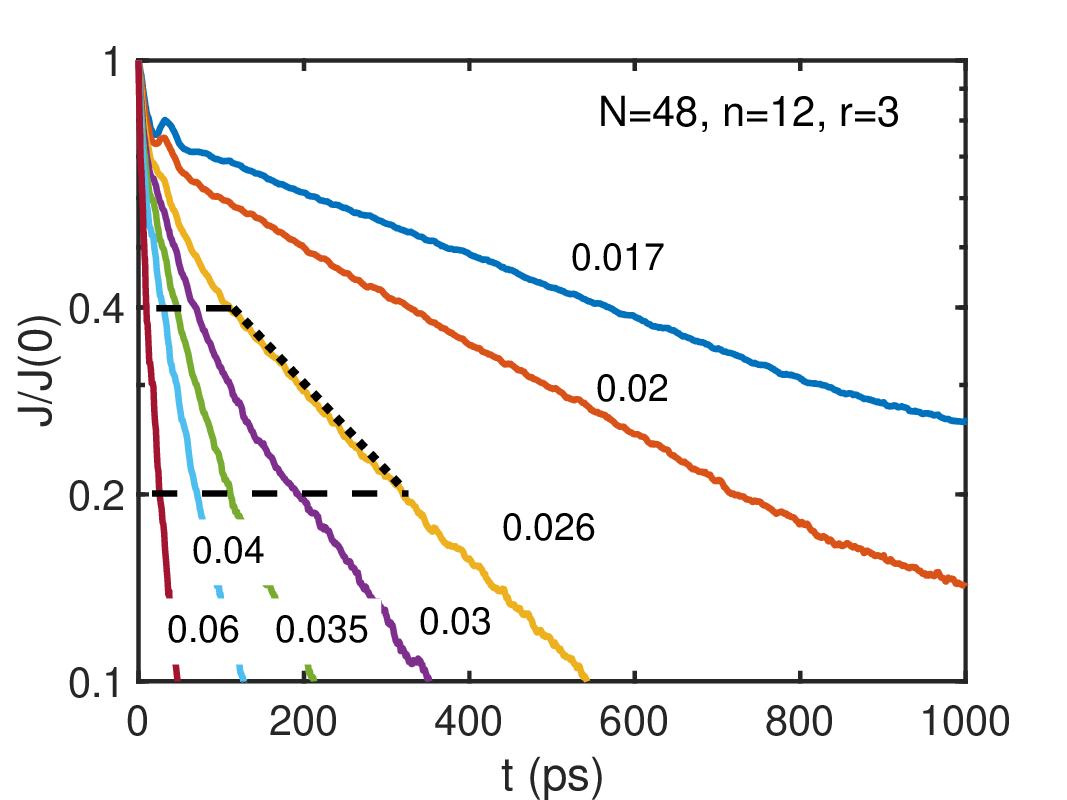} }}%
    \qquad
    \subfloat[\centering ]{{\includegraphics[width=7.5cm]{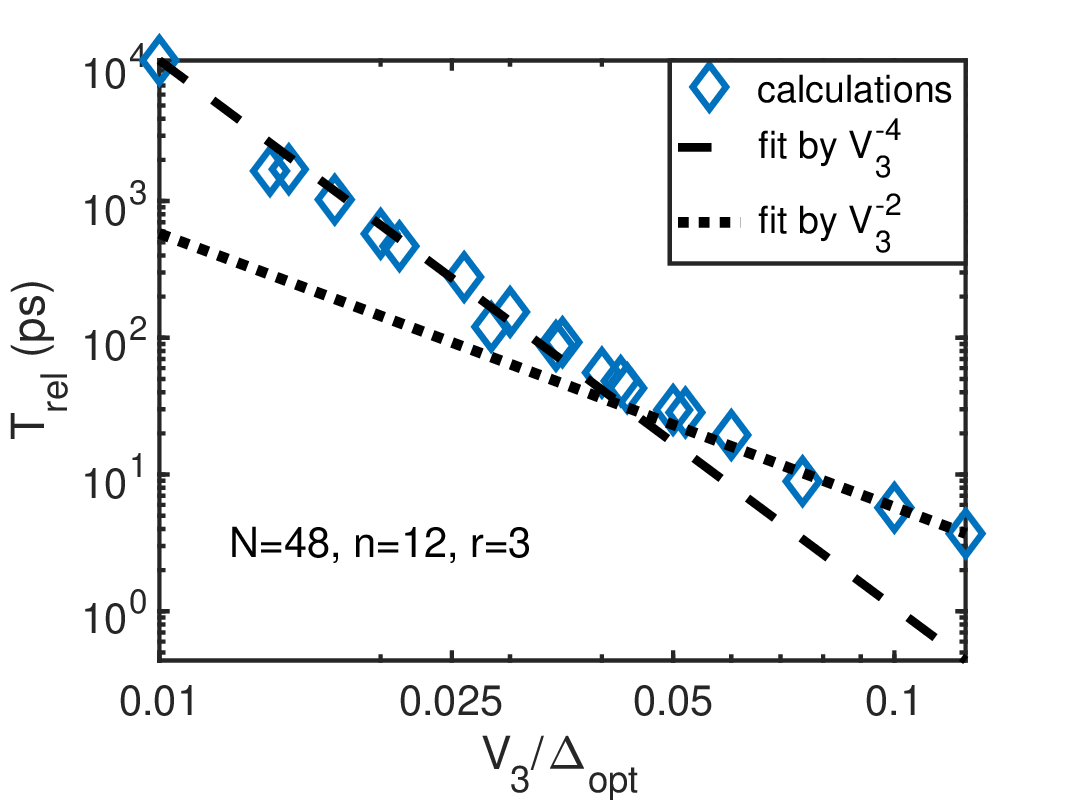} }}%
    \caption{(a) Time dependence of current for initially excited midband optical phonon   and different anharmonic coupling strengths $V_{3}/\Delta_{\rm opt}$ shown near each curve.  The negative inverse slope of the dashed line serves as an estimate for the relaxation time.  (b) Dependence of relaxation times on anharmonic coupling strengths fitted by $V_{3}^{-4}$ dependence for small couplings (dashed line) and  $V_{3}^{-2}$ dependence for large couplings (dotted line). In both graphs the bandwidth ratio is $r=3$. }%
    \label{fig:2StCurr-r3}%
\end{figure}

 First we report the results for the representative bandwidth ratio $r=3$ in Fig. \ref{fig:2StCurr-r3}. In Fig.  \ref{fig:2StCurr-r3}.a the time evolution of currents is shown for different anharmonic coupling constants. The current clearly relaxes in two stages and we focus on  the second stage. To extract the  relaxation time we fit the current time dependence by a single exponent for the part of relaxation occurring between $J/J(0)=0.4$ and $J/J(0)=0.2$ as shown by the dashed line. The relaxation time is estimated using the negative inverse slope of this line. The results are   sensitive to the fitting domain (use of the specific points $J/J(0)=0.4$ and $0.2$ to extract the relaxation rate), since relaxation is getting slower with time possibly reflecting its diffusive nature.  Our estimate gives a right  guess about the time of the transition  between ballistic and diffusive regimes. and parametric dependencies of this time nearly insensitive to its specific definition.  

The relaxation time  dependence on the anharmonic coupling strength is shown in Fig.  \ref{fig:2StCurr-r3}. At relatively small coupling $V_{3} \leq 0.05\Delta_{\rm opt}$ this dependence is perfectly consistent with the theory prediction of the inverse fourth power dependence Eq. (\ref{eq:scattFGREst}), while at large $V$ it switches to the weaker dependence probably because of  the failure of the Fermi's golden rule as a perturbation theory.  

\begin{figure}%
    \centering
    \subfloat[\centering ]{{\includegraphics[width=7.5cm]{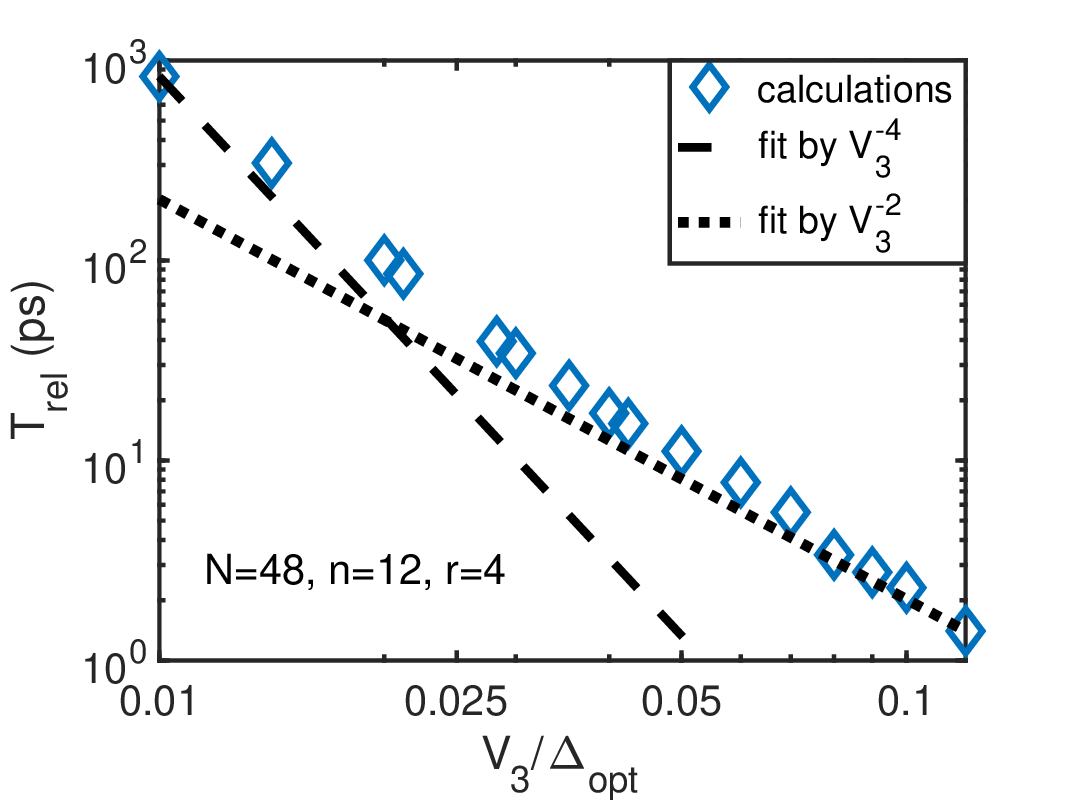} }}%
    \qquad
    \subfloat[\centering ]{{\includegraphics[width=7.5cm]{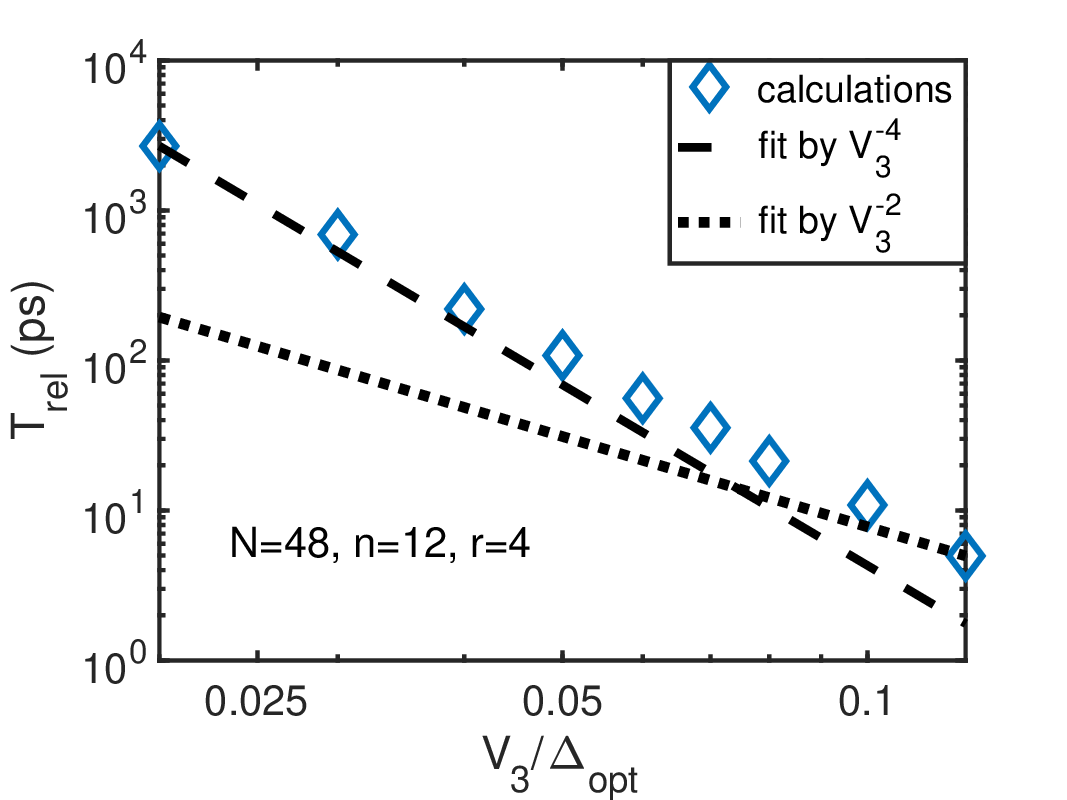} }}%
    \caption{ Dependence of relaxation times on anharmonic coupling strengths for $r=2$ (a) and $r=4$ (b) 
    fitted similarly to Fig. \ref{fig:2StCurr-r3}. }%
    \label{fig:2StCurr-r24}%
\end{figure}

Similar scaling of relaxation times is obtained for other two considered bandwidth  ratios $r=2$ and $r=4$ as illustrated in Fig. \ref{fig:2StCurr-r24}.a and b, respectively. The dependence $V_{3}^{-4}$ is seen in a wider domain for $r=4$ and in a very narrow domain for $r=2$ compared to $r=3$ that is the consequence of different separations from the threshold at $r=1$.

\begin{figure}
\includegraphics[scale=0.5]{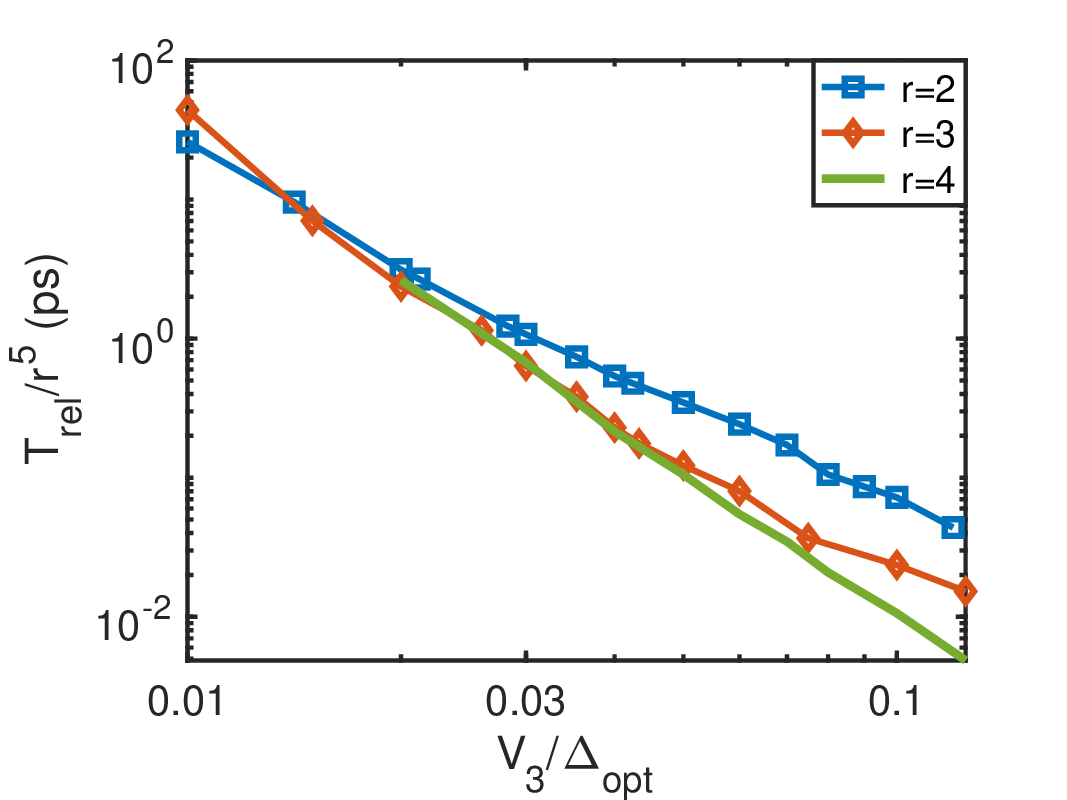} 
\caption{Dependence of rescaled relaxation times on anharmonic coupling for different bandwidth ratios.}
\label{fig:2Strdep}
\end{figure}

We also examine the dependence of relaxation times on the bandwidth ratio.  In our regime of interest $k_{\rm B}T=2\hbar\Delta_{\rm tr}$ the expected dependence can be expressed following Eq. (\ref{eq:scattFGREst}) as $T_{\rm rel} =r^5\Delta_{\rm opt}^3/(4CV_{3}^{4})$.  To check the relevance of this dependence we plot the rescaled relaxation times $T_{\rm rel}/r^{5}$ in Fig.  \ref{fig:2Strdep} for $r=2$, $3$ and $4$. The graphs for the bandwidth ratios $r=3$ and $4$ are perfectly consistent with the theory expectations. The  dependence for  $r=2$ deviates from those expectations.  This deviation is possibly because the case $r=2$ is close to the crossover regime $r=1$.  Two approaches are getting consistent  for $r=2$  at small $V_{3}$ if we use the resonant coupling $r$ dependence $V(k, -k, -k, k) \propto 1/(r^{2}-1)$ (see Eq. (\ref{eq:Period}))  to estimate the effective scattering matrix element replacing expected $r^{5}$ relaxation time dependence  with $r(r^{2}-1)^2$. 

Using the numerical estimate of the relaxation time  we can find   the dimensionless constant $C$  in Eq. (\ref{eq:scattFGREst}) as $2.7\cdot 10^{4}$.  This dimensionless numerical factor is huge, but this is not so surprising.  The typical transition matrix element $V(k, -k, -k, k)$ Eq. (\ref{eq:Period}) contains a large numerical factor of $\eta_{4}=16$.  At high  temperature it also acquires a factor expressing the number of thermal phonons $k_{\rm B}T/(\hbar \omega_{\rm tr}(k))$ estimated in Eq.  (\ref{eq:scattFGREst}) as  $k_{\rm B}T/(\hbar\Delta_{\rm tr})$.  However, the  actual energy $\hbar \omega_{\rm tr}(k)$ for the initial midband state $k=\pi/(2a)$ is a half of the bandwidth, so an extra factor of two naturally appears in the definition of the matrix element modifying  the numerical factor to $2\eta_{4} \sim 32$.  The squared matrix element within the Fermi's golden rule  acquires this numerical factor squared that is $32^2 \approx 1000$.

The remaining factor of $C/32^2 \approx 27$ in the definition of  the current relaxation rate can be originated from the integration over two wavevectors in the related Fermi's-golden rule expression for the current relaxation rate (see Refs.  \cite{pitaevskii2012physicalkinetics,
Busch94TransportCrossection}).    Another possible  origin for the extra rate  enhancement factor  of order of $10$ can be  due to a multistep current evolution involving first fast relaxation of the phonon to one of the states with a smaller energy with a subsequent backscattering from that state that can happen faster compared to that from the initial state due to larger number of participating  transverse phonons at lower energies.   Thus, the large numerical factor $C$ is quite reasonable for the current relaxation rate. 

\section{Acceleration of transport for a small initial  velocity of optical phonon}
\label{sec:exp}

We demonstrated in Secs.  \ref{sec:EnMomCons}, \ref{sec:num} that the  evolution of an optical phonon  due to its interaction  with  transverse phonons for the bandwidth ratio (transverse to optical ones) greater than unity can be separated into two stages.  In the first, fast  stage the phonon is scattered only forward, so its average velocity remains finite as for the ballistic transport.     Consequently, in the first stage we observed numerically the fast current reduction  to some finite value Fig. \ref{fig:CurClLt}.  This reduction is because  the initial midband  state used in Sec. \ref{sec:num} possesses the maximum velocity.  If the initial state is chosen near band edges, where the   velocity approaches minimum, then the forward scattering  to other faster states should increase the  velocity.  This is an interesting and untypical regime, where the relaxation enhances  the  current.  

Such   acceleration  can possibly explain the recently discovered  increase of the optical phonon  ballistic transport velocity  in alkane chains with the chain length \cite{2024IgorBallTr}.    The measurements carried out  similarly to the earlier work \cite{ab15ballistictranspexp} show that the phonon velocity  increases with increasing  the chain length from $14$\AA$/$ps for shorter chains  to around $48$\AA$/$ps for longer chains   containing more than $20$ CH$_{2}$ groups.   If the propagating wavepacket is initiated  at the top of a   band  where the group  velocity is small, it can be scattered towards the midband states, featuring much larger group velocities, still propagating in the forward direction.  Since  such a process became efficient at longer chains where the phonon has  sufficient time for scattering,  it would result in a speed increase.  Several optical bands of alkane chains can fit such conditions, including CH$_2$  wagging, CH$_2$ rocking, and C$-$C stretching bands \cite{ab21CompetChann}.  Then,  in long chains  the phonon gets re-scattered forward to midband states possessing much bigger group velocities up to $60$\AA$/$ps.  This redistribution emerges during the first stage of relaxation taking few picoseconds, while the second stage taking time orders of magnitude longer  is possibly not reached yet.  

\begin{figure}%
    \centering
    \subfloat[\centering ]{{\includegraphics[width=7.5cm]{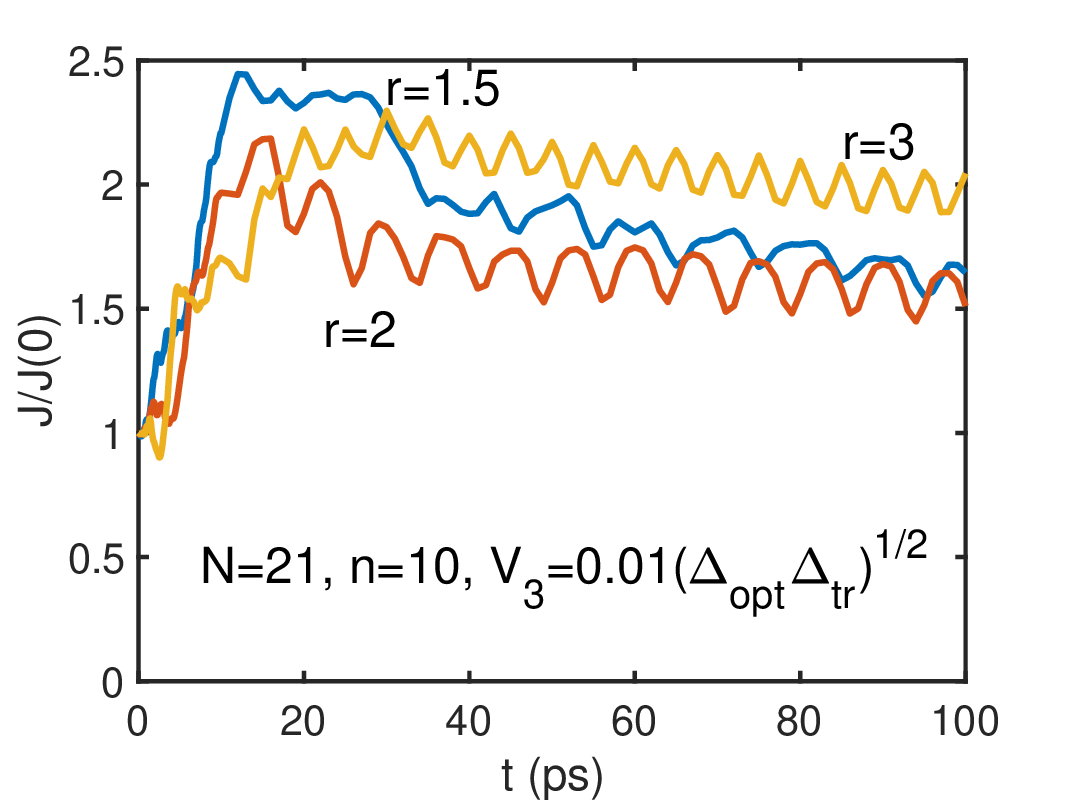} }}%
    \qquad
    \subfloat[\centering ]{{\includegraphics[width=7.5cm]{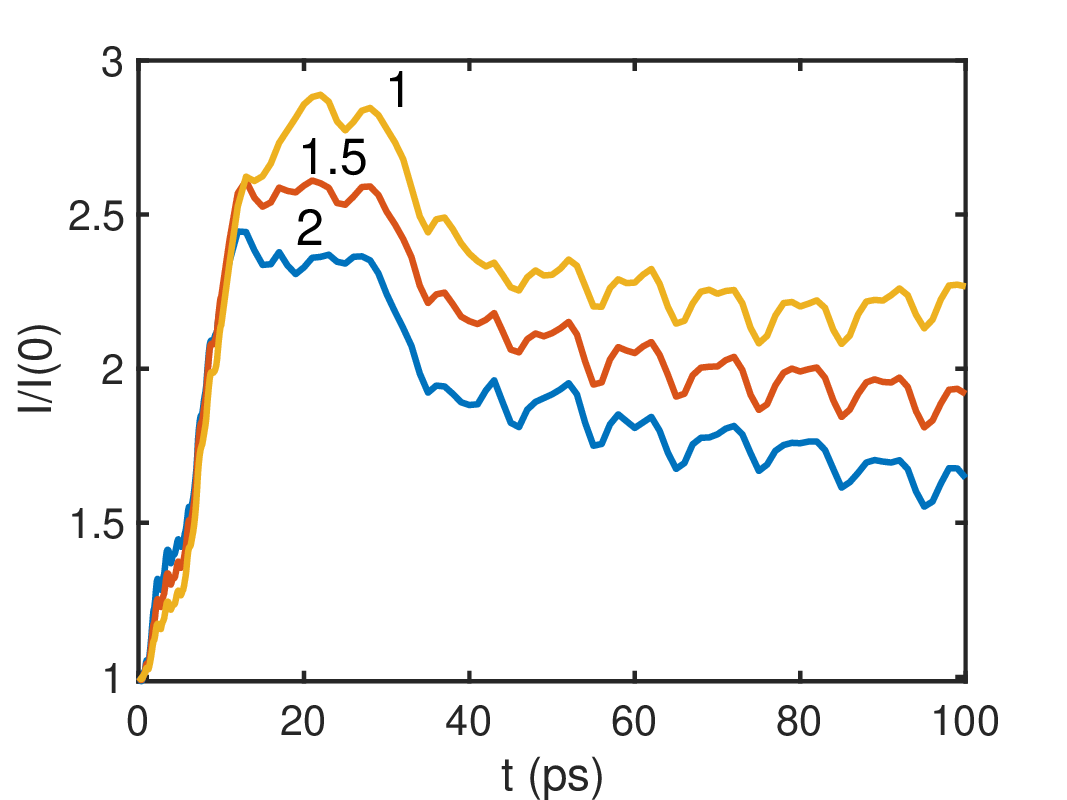} }}%
    \caption{(a) Raise of current due to anharmonic interaction for the initial state with a small group velocity ($N=21$, $n=10$) and  different bandwidth ratios $r$ shown for each graph.   The temperature is fixed at $k_{\rm B}T=2\hbar\Delta_{\rm tr}$(b)  Current for the bandwidth ratio  $r=1.5$ at different temperatures.  The ratio $k_{\rm B}T/(\hbar\Delta_{\rm tr})$ is shown above each graph.  }%
    \label{fig:Accel}%
\end{figure}

To demonstrate the optical phonon acceleration numerically we tried a variety of model parameters always choosing the initially excited optical phonon with a smallest (yet nonzero)  velocity  for different numbers of sites $N$ and anharmonic coupling strengths $V_{3}$.  The high temperature semiclassical regime ($k_{\rm B}T=2\hbar\Delta_{\rm tr}$) is considered since we observed a substantial increase of current only for $N>20$ sites, where  an accurate quantum mechanical treatment is problematic.  Below we report a substantial increase of current for  $N=21$ with the initial state wavevector $k=2\pi n/L$ and $n=10$ possessing the minimum group velocity.  

For our study we choose an  anharmonic coupling strengths  $V_3=0.01\sqrt{\Delta_{\rm opt}\Delta_{\rm tr}}$    for bandwidth ratios $1.5$, $2$, $3$ similarly to Sec.   \ref{sec:BacSchighT}.  The  absolute value of $V_{3}$ was chosen using guess and check method to maximize the current rise.   The time dependence of current for these specific parameters is shown in Fig. \ref{fig:Accel}.a.

According to Fig. \ref{fig:Accel}.a,  the current rises in the fast stage taking around $10$ ps by factor of $2$ or even $2.5$ depending on the specific bandwidth ratio.  The slow stage takes orders of magnitude longer, yet it limits the maximum current raise.  
The strongest increase of current is observed for a smallest considered   bandwidth ratio of $r=1.5$.

The verification of the proposed interpretation of the experimental data can be made modifying the temperature, which is the only controlling parameter that can be changed relatively easily and can be modeled using the present theory.  The effect of temperature is reported in Fig. \ref{fig:Accel}.B.  The rise of the current and, correspondingly, maximum transport velocity is clearly seen with the reduction of the temperature and it can be probed  experimentally to validate the present theory.    This rise with decreasing the temperature could be due to the suppression of slow backscattering processes limiting the raise of the current.  One should notice, however, that further reduction in temperature can reduce the transport rate due to phonon redistribution towards lower energies corresponding to their Boltzmann distribution.

\section{Conclusions}
\label{sec:concl}


We examined transport and decoherence of optical phonons in periodic chains due to their interaction with transverse acoustic phonons.  Similarly to the earlier work  \cite{ab2023Cherenkov} , where the interaction with longitudinal acoustic phonons was considered, we found two distinguishable dynamic  regimes depending on the relationship between acoustic and optical phonon bandwidths.  

In the typical regime of a narrower optical phonon band compared to acoustic phonon bands an optical phonon relaxation within the band emerges  in two stages.  The first stage takes several picoseconds.  It includes   fast equilibration  of the initially excited optical phonon within  the band states featuring similarly directed  group velocities by means of forward scattering accompanied by absorption or emission of transverse acoustic phonons forbidden for  longitudinal phonons  due to the Cherenkov's constraint \cite{ab2023Cherenkov}. 
Importantly,  this forward only scattering  supports the ballistic transport. 

If the  initial optical phonon velocity is smaller than its average velocity, then the phonon accelerates due to forward scattering to the states with a higher velocity.   The latter regime is possibly realized in the recent measurements of energy transport through alkane chains, where the optical phonon velocity increases with increasing the chain length \cite{2024IgorBallTr}.  

The second  stage of phonon relaxation involves its  backscattering. It converts the ballistic transport regime to diffusive but takes much longer time.  The backscattering occurs much slower compared to the forward scattering in the first stage, because for narrow optical bands  it requires higher  order anharmonic interactions.  

If the acoustic band is narrower (bandwidth ratio $r$ less than unity) then the optical phonon relaxes very quickly in about a few picoseconds from its initial state to all other states  within the band. This relaxation  leads to the substantial current reduction and changing the  transport from ballistic to  diffusive.  

Usually, acoustic phonon bands are wider compared to optical phonon bands \cite{ab15ballistictranspexp,ab2023Cherenkov}.  However,  our results for the opposite regime are also relevant for  other systems of interest including, for example,  electrons propagating  in  periodic molecules.  The electron energy band  can be broader than any phonon band, so the situation of $r<1$ is quite realistic.  

\begin{acknowledgements}

This work was supported by the National Science Foundation
(Grant No. CHE-2201027).

\end{acknowledgements}


\bibliography{Vibr}
\end{document}